\documentclass[aps,prl,twocolumn,nopacs,superscriptaddress]{revtex4}

\usepackage{graphicx}  
\usepackage{dcolumn}   
\usepackage{amssymb}   
\usepackage{amsmath}
\usepackage{units}
\usepackage[ansinew]{inputenc}
\usepackage[dvipsnames]{xcolor}

\usepackage{hyperref}

\usepackage[type1]{libertine}                                        
\usepackage{textcomp}
\usepackage[scaled=.85]{beramono}
\usepackage[libertine,cmintegrals,cmbraces,vvarbb,slantedGreek]{newtxmath}
\usepackage[scr=boondoxo]{mathalfa}
\usepackage{bm}
\usepackage[lf]{carlito}

\usepackage{braket}

\makeatletter
\DeclareRobustCommand{\cev}[1]{%
  \mathpalette\do@cev{#1}%
}
\newcommand{\do@cev}[2]{%
  \fix@cev{#1}{+}%
  \reflectbox{$\m@th#1\vec{\reflectbox{$\fix@cev{#1}{-}\m@th#1#2\fix@cev{#1}{+}$}}$}%
  \fix@cev{#1}{-}%
}
\newcommand{\fix@cev}[2]{%
  \ifx#1\displaystyle
    \mkern#23mu
  \else
    \ifx#1\textstyle
      \mkern#23mu
    \else
      \ifx#1\scriptstyle
        \mkern#22mu
      \else
        \mkern#22mu
      \fi
    \fi
  \fi
}
\makeatother

\renewcommand{\Re}{\text{Re}}

\begin{document}

\title{Magnetic Impurities on Superconducting Surfaces: Phase Transitions and the Role of Impurity-Substrate Hybridization}

\author{Haonan Huang}
\affiliation{Max-Planck-Institut f\"ur Festk\"orperforschung, Heisenbergstraße 1,
70569 Stuttgart, Germany}
\author{Robert Drost}
\affiliation{Max-Planck-Institut f\"ur Festk\"orperforschung, Heisenbergstraße 1,
70569 Stuttgart, Germany}
\author{Jacob Senkpiel}
\affiliation{Max-Planck-Institut f\"ur Festk\"orperforschung, Heisenbergstraße 1,
70569 Stuttgart, Germany}
\author{Ciprian Padurariu}
\affiliation{Institut für Komplexe Quantensysteme and IQST, Universität Ulm, Albert-Einstein-Allee 11, 89069 Ulm, Germany}
\author{Björn Kubala}
\affiliation{Institut für Komplexe Quantensysteme and IQST, Universität Ulm, Albert-Einstein-Allee 11, 89069 Ulm, Germany}
\author{Alfredo Levy Yeyati}
\affiliation{Departamento de F\'{\i}sica Te\'orica de la Materia Condensada and
Condensed Matter Physics Center (IFIMAC), Universidad Aut\'onoma de Madrid, 28049 Madrid, Spain}
\author{Juan Carlos Cuevas}
\affiliation{Departamento de F\'{\i}sica Te\'orica de la Materia Condensada and
Condensed Matter Physics Center (IFIMAC), Universidad Aut\'onoma de Madrid, 28049 Madrid, Spain}
\author{Joachim Ankerhold}
\affiliation{Institut für Komplexe Quantensysteme and IQST, Universität Ulm, Albert-Einstein-Allee 11, 89069 Ulm, Germany}
\author{Klaus Kern}
\affiliation{Max-Planck-Institut f\"ur Festk\"orperforschung, Heisenbergstraße 1,
70569 Stuttgart, Germany}
\affiliation{Institut de Physique, Ecole Polytechnique Fédérale de Lausanne, 1015 Lausanne, Switzerland}
\author{Christian R. Ast}
\email[Corresponding author; electronic address:\ ]{c.ast@fkf.mpg.de}
\affiliation{Max-Planck-Institut f\"ur Festk\"orperforschung, Heisenbergstraße 1,
70569 Stuttgart, Germany}

\date{\today}

\begin{abstract}
Spin-dependent scattering from magnetic impurities inside a superconductor gives rise to Yu-Shiba-Rusinov (YSR) states within the superconducting gap. As such, YSR states have been very successfully modeled with an effective scattering potential (Kondo impurity model). Using a scanning tunneling microscope, we exploit the proximity of the tip to a surface impurity and its influence on the YSR state to make a quantitative connection between the YSR state energy and the impurity-substrate hybridization. We corroborate the coupling between impurity and substrate as a key energy scale for surface derived YSR states using the Anderson impurity model in the mean field approximation, which accurately explains our observations. The model allows to decide on which side of the quantum phase transition the system resides based on additional conductance measurements. We propose that the Anderson impurity model is much more appropriate to describe YSR states from impurities \textit{on} a superconducting surface than the Kondo impurity model, which is more appropriate for impurities \textit{inside} a superconductor. We thus provide a first step towards a more quantitative comparison of experimental data with fully correlated calculations based on the Anderson impurity model.
\end{abstract}

\maketitle

\section{Introduction}

\begin{figure}
\centerline{\includegraphics[width = \columnwidth]{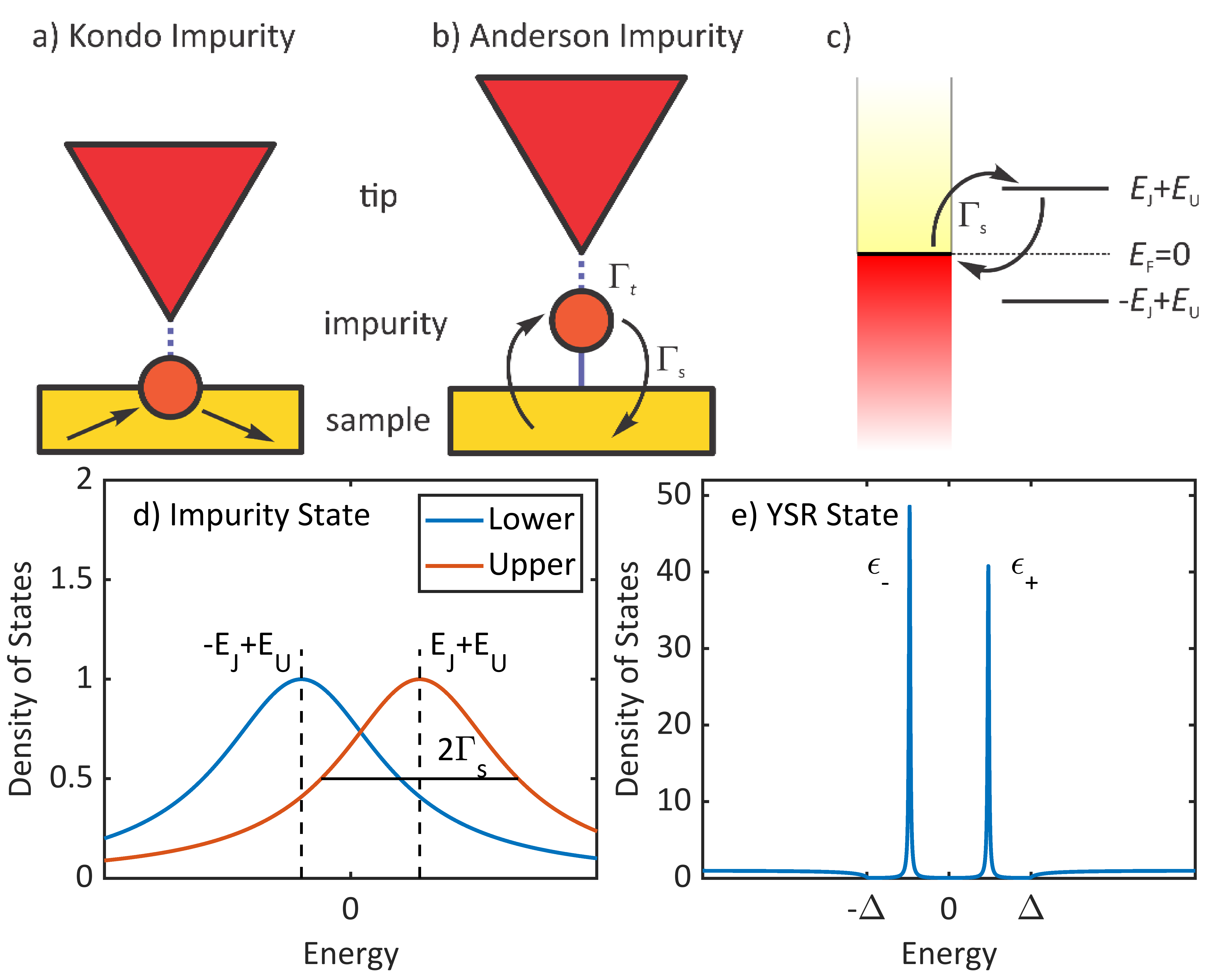}}
\caption{a) In the Kondo impurity model, the YSR state arises due to scattering from a spin-dependent impurity potential. (b) In the Anderson impurity model, the YSR state arises due to hopping to and from an impurity state. (c) Energy diagram of the Anderson impurity model. The coupled impurity features an occupied state below the Fermi level at $-E_\text{J}+E_\text{U}$ and an unoccupied state above the Fermi level at $E_\text{J}+E_\text{U}$. The coupling strength is given by $\Gamma_\text{s}$. (d) Spectral functions of the two Anderson impurity states in the normal conducting state. There is significant overlap between these two states. (e) The resulting YSR states in the superconducting regime. Note the difference in energy scale between (d) and (e).} \label{fig:intro}
\end{figure}

The impurity problem is one of the most extensively studied phenomena in condensed matter physics because it not only caters to fundamental interest in the local perturbation of a host material, but also has technological relevance in the design of specific properties through doping. The impact of impurities on the host material is broad ranging from having no effect for weak non-magnetic impurities in an $s$-wave superconductor (Anderson theorem) \cite{anderson_theory_1959,abrikosov_superconducting_1959} to creating complex many-body interactions between a magnetic impurity in a normal conducting host (Kondo effect) \cite{kondo_resistance_1964}. Somewhere in between, we find the so-called Yu-Shiba-Rusinov (YSR) states \cite{yu_bound_1965,shiba_classical_1968,rusinov_superconductivity_1969}, which arise from magnetic impurities in a superconducting host. YSR states have been quite successfully modelled as a combination of spin-dependent and spin-independent scattering potentials within the Kondo impurity model (see Fig.\ \ref{fig:intro}(a)) \cite{salkola_spectral_1997,flatte_local_1997,flatte_local_1997-1}. As such, this YSR model provides a simple and straightforward framework that has gone quite far in explaining numerous observations.

It is obvious that surface adsorbed impurities have more spatial degrees of freedom to relax when hybridizing with the host than bulk impurities. Impurity-substrate hybridization, however, is only implicitly contained in the Kondo impurity model \cite{schrieffer_relation_1966}. A more detailed description is offered by the largely equivalent, albeit more general, Anderson impurity model (see Fig.\ \ref{fig:intro}(b)). It explicitly introduces an impurity-substrate hybridization parameter $\Gamma_\text{s}$, which plays a key role for the adsorption of impurities at surfaces. The Anderson impurity model also has the added benefit that it encompasses the Kondo effect as well as Andreev bound states, into which YSR states are embedded in a more general context \cite{zitko_effects_2011,zitko_shiba_2015,kadlecova_quantum_2017,kadlecova_practical_2019}. In fact, this model provides a benchmark for the analysis of Josephson and Andreev transport through quantum dots (for a review see \cite{martin-rodero_josephson_2011}). Also, as tunneling is often understood as going \textit{through} the impurity (i.\ e.\ the YSR state), the impurity-substrate coupling will influence the conductance as well, which can be modeled much better within the Anderson impurity model \cite{cuevas_molecular_2010}. In order to ascertain these relations, a quantitative connection between the impurity-substrate hybridization and the behavior of the YSR state is needed.

Here, we show that the binding energy of YSR states for surface adsorbed impurities does not just depend on the magnetic and non-magnetic properties of the impurity, but also largely depends on the coupling between the impurity and the substrate. We use ultralow temperature scanning tunneling microscopy (at 10\,mK) to probe YSR states in intrinsic surface impurities on a superconducting V(100) substrate with a superconducting vanadium tip. Approaching the tip to an impurity with YSR state induces an interaction between the tip and the impurity (e.\ g.\ attractive force \cite{ternes_force_2008,ternes_interplay_2011}), which manifests itself as a change in the binding energy of the YSR state. This has been observed in a number of systems giving the phenomenon a more general character \cite{ternes_scanning_2006,brand_electron_2018,farinacci_tuning_2018,malavolti_tunable_2018,kezilebieke_interplay_2018}. However, both a decrease \cite{farinacci_tuning_2018} as well as an increase \cite{malavolti_tunable_2018} in impurity-substrate coupling has been found. Using the Anderson impurity model in the mean field approximation, we are able to quantify the relation between the change in the YSR state binding energy and the impurity-substrate coupling. We independently confirm the change in the impurity-substrate coupling through the distance dependence of the normal state conductance.

Further, we use this connection between impurity-substrate coupling and normal state conductance to determine, whether the YSR state is in the weak or strong scattering regime. For weak impurity-substrate coupling, the spin-dependent scattering potential will be weak and the impurity spin will be unscreened. As the impurity-substrate coupling increases, the system undergoes a quantum phase transition to a screened impurity spin in a strong scattering potential \cite{salkola_spectral_1997,flatte_local_1997,flatte_local_1997-1}. We demonstrate how to apply this model to determine on which side of the quantum phase transition the system is, which is \textit{a priori} impossible to judge from the tunneling spectrum alone due to the symmetry of the YSR state energies in the spectral function.

\section{Experiment}

We prepare single crystal V(100) surfaces through cycles of sputtering and annealing (700$^\circ$C). Due to the intrinsic presence of oxygen in the bulk (99.8\% purity) and aggregation to the surface during annealing, the surface features a ($5\times1$) reconstructed oxygen layer. The most abundant impurities visible in STM topography image are most likely oxygen vacancies, while carbon is also expected to have a non-negligible concentration which, however, is not directly visible. Some of the oxygen vacancies in a certain chemical environment, feature single and well defined intrinsic YSR states. Due to the complexity of the surface and various possibilities of the internal structure of the impurity, the YSR states show wide-spread energy distribution and different response to tip approach \cite{rodrigo_use_2004,guillamon_scanning_2008,supinf}. The experiments have been performed in ultra high vacuum and at a base temperature of 10\,mK. The gap parameter of vanadium in the sample as well as in the tip is $\Delta_\text{s} = \Delta_\text{t} = 760\,\upmu$eV unless otherwise noted.

\section{Results}

\subsection{Distance dependence of YSR States}

\begin{figure}
\centerline{\includegraphics[width = \columnwidth]{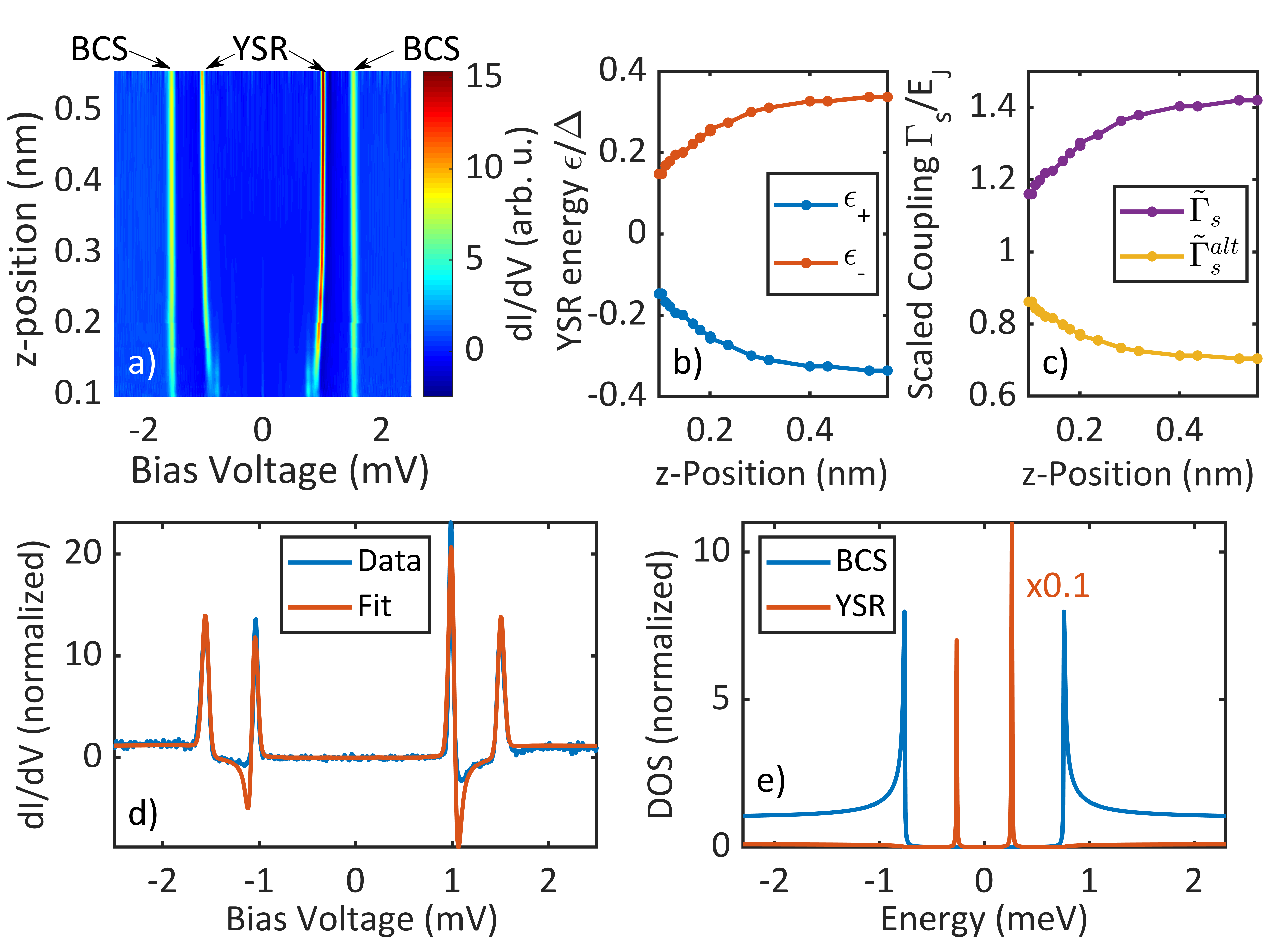}}
\caption{(a) Series of normalized differential conductance spectra through an impurity with YSR states measured with a superconducting tip as function of tip-sample distance (z-position). The YSR states move, while the coherence peaks (BCS) do not. At closer distances (bottom), higher order phenomena (Josephson effect at zero voltage and multiple Andreev processes near the YSR states) are visible. (b) Extracted YSR state energies as function of tip-sample distance. (c) Scaled coupling parameter $\tilde{\Gamma}_{s}$ calculated from (b). The values for  $\tilde{\Gamma}_{s}^\text{alt}$ have been calculated by exchanging $\varepsilon_+\leftrightarrow\varepsilon_-$ (for details see text). (d) Fit of a differential conductance spectrum at low conductance, where higher order processes are suppressed. We fit two channels, one of which probes the YSR state and the other probes an empty gap. (e) Density of states of the YSR state and the empty BCS gap as used in the fit in (d).} \label{fig:plot2}
\end{figure}

Some of the YSR states originating from the intrinsic impurities at the V(100) surface change their energy as a function of tip-sample distance. One example is shown in Fig.\ \ref{fig:plot2}. In panel (a), a series of differential conductance spectra is shown as a function of tip-sample distance ($z$-position). A single pair of YSR states can be identified inside the gap (marked by YSR arrows), which changes its energy position as a function of $z$-position. The observation of coherence peaks (marked by BCS arrows) at $\Delta_\text{t} + \Delta_\text{s}$ is an indication that there is a second transport channel not featuring a YSR state inside the gap, which will be discussed in more detail below. The energies at which the YSR states are observed have been extracted and plotted as a function of tip sample distance in Fig.\ \ref{fig:plot2}(b).

The YSR state at positive (negative) bias voltage has been plotted in red (blue). As we will show below, the YSR states are in the strong spin-dependent scattering limit beyond the quantum phase transition \cite{salkola_spectral_1997}. In that regime, that branch of YSR states with positive values in the weak scattering limit ($\epsilon_+>0$, called positive branch in the following) has moved to negative energies, $\epsilon_+<0$, as shown in Fig.\ \ref{fig:plot2}(b). \textit{A priori}, however, it is not possible to say on which side of the quantum phase transition the YSR state is in each case. A more detailed analysis of the YSR state properties as function of tip-sample distance is necessary. For this, we have acquired different spectra along the $z$-axis (tip-sample distance) and over a distance of about 470\,pm, which corresponds to a change in tunneling current of about four orders of magnitude. Yet, we have stayed mostly in the tunneling regime (see below). Only in the last part, we find total transmissions $\tau>0.1$, where higher order processes are observed and the opening of new transport channels becomes more likely.

\subsection{Distance Dependence of the Conductance}

\begin{figure}
\centerline{\includegraphics[width = \columnwidth]{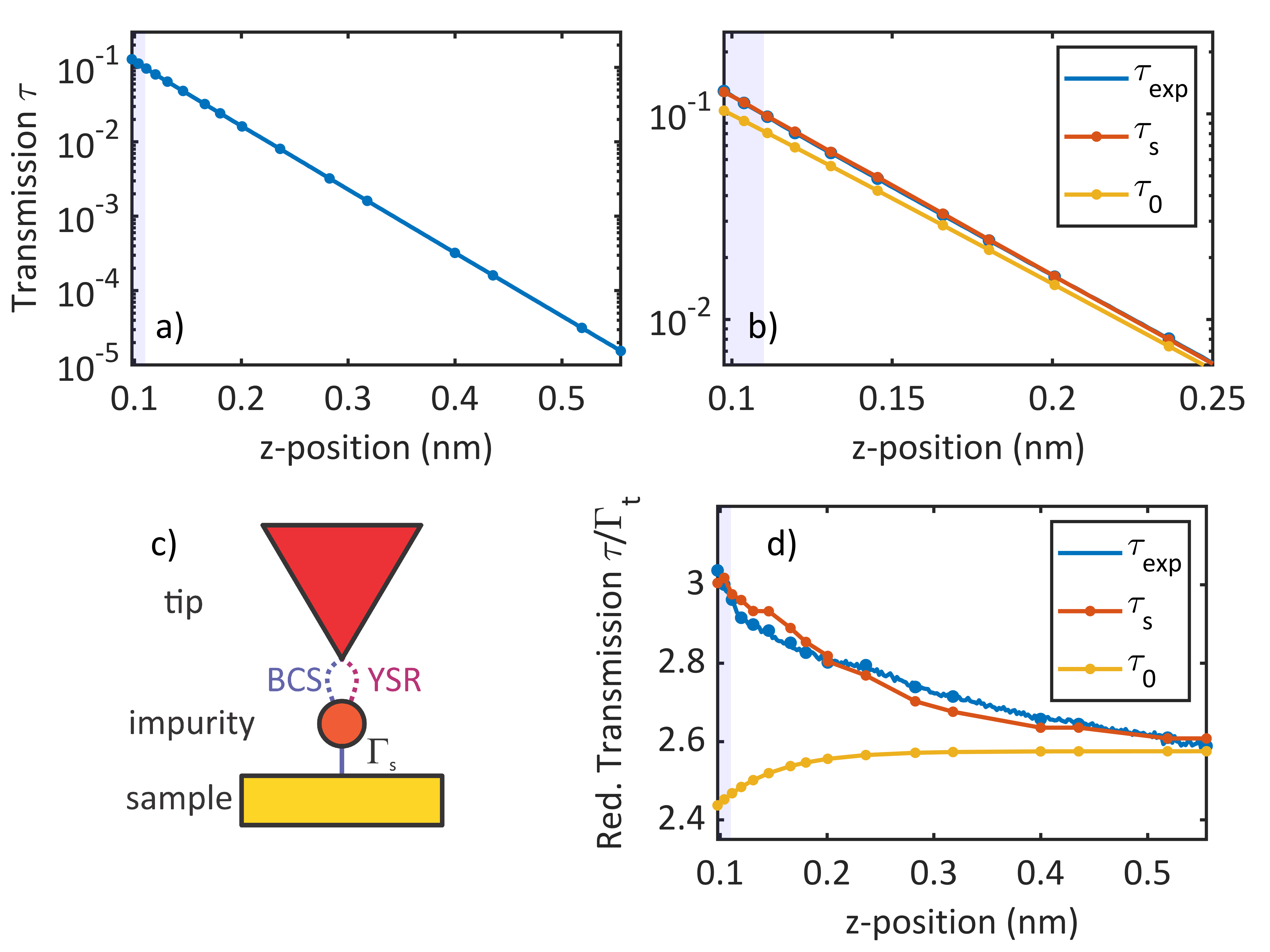}}
\caption{(a) Experimental normal state transmission $\tau_\text{exp}=G_\text{N}/G_0$ as function of tip-sample distance. The exponential dependence is clearly visible. (b) Zoom-in, where the tip is close to the sample. The measured transmission $\tau_\text{exp}$ (blue) is compared to a transmission assuming an impurity-substrate coupling that is constant ($\tau_0$, yellow) and one that changes like $\tilde{\Gamma}_\text{s}$ in Fig.\ \ref{fig:plot2}(c) ($\tau_\text{s}$, red). (c) The tip probes two channels to the impurity, one of which is leading through the YSR state and the other through an empty BCS gap. Both channels couple to the substrate through the same impurity-substrate channel(s) $\Gamma_\text{s}$. (d) Reduced transmission $\tau_\text{exp}/\tilde{\Gamma}_\text{t}$ with the data labeled as in (b).} \label{fig:analysis2}
\end{figure}

The normal state conductance is extracted from the differential conductance measured at a bias voltage much larger than $\Delta_\text{t} + \Delta_\text{s}$. The normalized normal state conductance (transmission) $\tau = G_\text{N}/G_0$ ($G_\text{N}$: normal state conductance; $G_0=2e^2/h$: quantum of conductance; $e$: elementary charge; $h$: Planck constant) corresponding to the data set in Fig.\ \ref{fig:plot2}(a) is shown in Fig.\ \ref{fig:analysis2}(a) as a function of tip-sample distance ($z$-position). Its behavior is dominated by the exponential increase in the tunnel coupling between tip and impurity. However, we will show below that there are deviations from the exponential behavior, which are related to the changes in the impurity-substrate coupling. The transport current through the impurity does not only depend on the tunnel coupling between the tip and the impurity, but also on the coupling between the impurity and the substrate \cite{cuevas_molecular_2010}. These deviations nicely explain the changes in the YSR state energies. The Anderson model in the mean field approximation is ideally suited to provide a unified description of these observations. Since the impurity-substrate coupling is an explicit parameter, we can establish a direct relation between the YSR state energy and the changes in the normal state conductance. This is not easy to do within the Kondo impurity model.

\subsection{Anderson Impurity Model}

The Anderson impurity model has been successfully applied to a number of impurity problems involving magnetic as well as non-magnetic impurities \cite{anderson_localized_1961}. It allows correlation effects to be taken into account to different degrees of complexity \cite{yoshioka_numerical_2000,zitko_shiba_2015}. For the case that we consider here, where the Kondo temperature is typically smaller than the superconducting gap, a mean field approximation becomes appropriate, as shown in Refs. \onlinecite{vecino_josephson_2003,martin-rodero_andreev_2012}.

A schematic energy diagram is shown in Fig.\ \ref{fig:intro}(c). The system is described by the superconducting substrate (left) and the impurity having one occupied level at $-E_\text{J}+E_\text{U}$ and one unoccupied energy level at $E_\text{J}+E_\text{U}$ (right), which are coupled to the substrate by a hopping parameter $\Gamma_\text{s}$. The energy $E_\text{J}$ describes an effective Zeeman splitting and $E_\text{U}$ is an energy shift accounting for particle-hole asymmetry ($E_\text{U}=0$ implies particle-hole symmetry). Here, we restrict ourselves to using the energies $E_\text{J}$ and $E_\text{U}$ as fit parameters, keeping in mind that a selfconsistent treatment of the spin density of states may provide more insight on the origin of the magnetic properties as well as spin fluctuations in the impurity on the substrate.

The Green's function of the impurity in the mean field Anderson impurity model can be straightforwardly written in $2\times2$ Nambu space as
\begin{equation}
    G_\text{I}(\omega) = \left[\omega\sigma_0 - E_\text{J}\sigma_0 + E_\text{U}\sigma_3 - \Gamma_\text{s}\sigma_3g_\text{sc}(\omega)\sigma_3\right]^{-1},
    \label{eq:gimp}
\end{equation}
where $\sigma_i$ are the Pauli matrices. We assume that the coupling between tip and impurity is much smaller than the coupling between impurity and sample, i.\ e.\ $\tau\ll 1$, such that we can ignore it in this calculation. Further, $g_\text{sc}(\omega)$ is the dimensionless Green's function of the superconducting substrate (normalized to the density of states) with
\begin{equation}
    g_\text{sc}(\omega) = \frac{(\omega +  i\gamma)\sigma_0 - \Delta_\text{s}\sigma_1}{\sqrt{\Delta_\text{s}^2 - (\omega +  i\gamma)^2}},
\end{equation}
where $\Delta_\text{s}$ is the order parameter of the substrate and $\gamma$ is a phenomenological broadening parameter (cf.\ Dynes \textit{et al.}\ \cite{dynes_direct_1978}). For more details, refer to the Supporting Information \cite{supinf}.

The spectral function $A(\omega) = -\text{Im}\text{Tr'}G(\omega)$ of Eq.\ \eqref{eq:gimp} features two impurity states at $-E_\text{J}+E_\text{U}$ and $E_\text{J}+E_\text{U}$ each having a width $2\Gamma_\text{s}$ (cf.\ Fig.\ \ref{fig:intro}(d) for the normal conducting state) along with a superconducting gap having an order parameter $\Delta_\text{s}$ and possibly extremely sharp pairs of subgap states depending on the relation between the parameters $E_\text{J}$, $E_\text{U}$, $\Gamma_\text{s}$, and $\Delta_\text{s}$ (cf.\ Fig.\ \ref{fig:intro}(e) for typical YSR states inside the superconducting gap). Here, Tr' denotes the trace with a change in sign for the energy axis in the hole part of the Green's function.

For the purpose of analyzing the above data, we reduce the generality of Eq.\ \eqref{eq:gimp} by assuming strong impurity-substrate coupling, i.\ e.\ $\Gamma_\text{s}\gg\Delta_\text{s}$. This assumption generally holds for surface adsorbed impurities and reflects the conditions, in which the YSR states within the Kondo impurity model are described. The resulting Green's function is
\begin{equation}
    G(\omega) = \frac{\Gamma_\text{s}\omega\sigma_0 + (E_\text{J}\sigma_0+E_\text{U}\sigma_3)\sqrt{\Delta_\text{s}^2-\omega^2} + \Gamma_\text{s}\Delta_\text{s}\sigma_1}{2E_\text{J}\Gamma_\text{s}\omega - (\Gamma_\text{s}^2-E_\text{J}^2+E_\text{U}^2)\sqrt{\Delta_\text{s}^2-\omega^2}},
    \label{eq:GYSR}
\end{equation}
without broadening parameter, which can be included by $\omega \rightarrow \omega +  i\gamma$.
The energies $\varepsilon_\pm$ of the YSR states are located, where $G(\omega)$ becomes singular:
\begin{equation}
    \varepsilon_\pm = \pm\Delta_\text{s}\frac{E_\text{J}^2-\Gamma_\text{s}^2-E_\text{U}^2}
    {\sqrt{\left(\Gamma_\text{s}^2+(E_\text{J}-E_\text{U})^2\right)\left(\Gamma_\text{s}^2+(E_\text{J}+E_\text{U})^2\right)}},
    \label{eq:ysrenergy}
\end{equation}
which has a very similar structure as the result from the Kondo impurity model. The similarity becomes even more obvious when simplifying Eq.\ \eqref{eq:ysrenergy} by assuming particle-hole symmetry, i.\ e.\ $E_\text{U}=0$,
\begin{equation}
    \varepsilon_\pm = \underbrace{\pm\Delta_\text{s}\frac{E_\text{J}^2-\Gamma_\text{s}^2}{E_\text{J}^2+\Gamma_\text{s}^2}}_\text{Anderson} = \underbrace{\pm\Delta_\text{s}\frac{1-J^2}{1+J^2_{\text{\textcolor{white}{J}}}}}_\text{Kondo}.
    \label{eq:ysrenergysym}
\end{equation}
The parameter $J$ is the spin-dependent scattering potential in the Kondo impurity model with $J = \frac{1}{2}n_0 j s$ in the classical limit, where $n_0$ is the density of states in the substrate, $j$ is the exchange coupling, and $s$ is the impurity spin \cite{salkola_spectral_1997,flatte_local_1997,flatte_local_1997-1}. The parameters describing the YSR states in the Anderson impurity model and the Kondo impurity model are related through the Schrieffer-Wolff-like transformation in the strong coupling limit (cf.\ \cite{schrieffer_relation_1966,zitko_shiba_2015}).
\begin{equation}
    J = \frac{\Gamma_\text{s}}{E_\text{J}}.
\end{equation}

For the following data analysis, we assume that the parameters which are more related to the intrinsic properties of the impurity $E_\text{J}$ and $E_\text{U}$ are constant as function of tip-sample distance, while the impurity-substrate coupling $\Gamma_\text{s}$ can vary. This is a sensible assumption of some generality, which has been used before in a somewhat different context for YSR states in molecules adsorbed on a superconducting surface \cite{bauer_microscopic_2013,kezilebieke_interplay_2018}. However, we have to keep in mind that in a selfconsistent treatment $E_\text{J}$ becomes a function of $\Gamma_\text{s}$, which may lead to small corrections. We also assume particle-hole symmetry (i.\ e.\ $E_\text{U}=0$), which is justified because we can show that $E_\text{U}$ is small compared to $\Gamma_\text{s}$ \cite{supinf}. Using the branch $\varepsilon_+$, we find for the impurity-substrate coupling
\begin{equation}
    \Gamma_\text{s} = E_\text{J}\frac{\sqrt{1-\varepsilon_+/\Delta_\text{s}}}{\sqrt{1+\varepsilon_+/\Delta_\text{s}}}.
    \label{eq:ysrcoupling}
\end{equation}
The symmetry of the YSR state energies makes it \textit{a priori} impossible to decide, on which side of the quantum phase transition the system is, i.\ e.\ if $\Gamma_s < E_J$ or $\Gamma_s > E_J$. Therefore, aside from the coupling $\Gamma_\text{s}$, we calculate an alternative coupling $\Gamma_\text{s}^\text{alt}$ by exchanging the values $\varepsilon_+\leftrightarrow\varepsilon_-$, which changes effectively from one side to the other side of the quantum phase transition.

Using Eq.\ \eqref{eq:ysrcoupling}, we calculate the distance dependent coupling $\Gamma_\text{s}$ and $\Gamma_\text{s}^\text{alt}$. The results are shown in Fig.\ \ref{fig:plot2}(c) in units of $E_\text{J}$, for which we define the scaled coupling $\tilde{\Gamma}_\text{s}=\Gamma_\text{s}/E_\text{J}$ and $\tilde{\Gamma}_\text{s}^\text{alt}=\Gamma_\text{s}^\text{alt}/E_\text{J}$. We can see directly, that for the $\tilde{\Gamma}_\text{s}$ branch the coupling reduces as the tip-sample distance reduces. Such a behavior can be expected, if attractive forces from the tip pull the impurity away from the substrate in the tunneling regime \cite{ternes_interplay_2011,farinacci_tuning_2018}. However, concomitant circumstances, e.\ g.\ changes in the local density of states, may just as well result in an increase in coupling, yielding the behavior described by the $\tilde{\Gamma}_\text{s}^\text{alt}$ data \cite{cuevas_evolution_1998,malavolti_tunable_2018}. We will directly address this point below by implementing a model to link the extracted impurity-substrate coupling to the measured normal state conductance.

In the following, we will show that analyzing the evolution of both the impurity-substrate coupling $\Gamma_\text{s}$ and the normal state conductance $G_\text{N}$ as function of tip-sample distance $z$, we are able to determine, on which side of the quantum phase transition the system is.

\subsection{Distance dependence of the Impurity-Substrate Coupling}

The normal state conductance $G_\text{N}$ of the junction not only depends on the tunneling between tip and impurity but also on the coupling between impurity and substrate. The latter may change when the distance $z$ between tip and impurity is tuned due to attractive or repulsive forces between tip and impurity. In addition, an understanding of the distance dependence $G_\text{N}(z)$ requires an analysis of possible transport channels involved, which we discuss in the following.

As can be seen in Fig.\ \ref{fig:intro}(e), YSR states alone give rise to two distinct peaks in the density of states completely quenching the coherence peaks. This is in contrast to our experimental observations depicted in Fig.\ \ref{fig:plot2}(d), where two additional peaks appear at $\pm(\Delta_\text{t}+\Delta_\text{s})$ as coherence peaks in the spectrum. We conclude that we have to assume two transport channels, which we assume to be independent. Microscopically, we envision these two channels as coming from two different orbitals, one of which features a YSR state due to the interaction with the substrate and the other does not (cf.\ Fig.\ref{fig:analysis2}(c)).

Accordingly, we calculate the total normal state conductance $G_\text{N}$ as the sum of the two contributions (assuming that $E_\text{U}=0$)
\begin{equation}
    G_\text{N} = G_\text{YSR} + G_\text{BCS} = p \frac{4\Gamma_\text{s}\Gamma_\text{t}}{(\Gamma_\text{s}+\Gamma_\text{t})^2+E_\text{J}^2}G_0 + (1-p)\frac{4\Gamma_\text{s}\Gamma_\text{t}}{(\Gamma_\text{s}+\Gamma_\text{t})^2}G_0,
    \label{eq:transmission}
\end{equation}
where $p$ is the relative signal contributions and $\Gamma_\text{t}= \Gamma_{\text{t}0} \text{exp}[-(z-z_0)/z_1]$ is the exponentially varying tunnel coupling between the tip and the impurity (for details see \cite{supinf}). The parameters $\Gamma_{\text{t}0}$ and $z_1$ are the only fit parameters to model the normal state conductance, while $z_0$ just represents the arbitrary position of the origin of the $z$-axis. The two fit parameters can be determined in the regime, where the tip is far away from the sample, such that the influence on the impurity-substrate coupling is smallest. We further assume that the two different channels use the same impurity-substrate channel(s), which is illustrated in Fig.\ \ref{fig:analysis2}(c). Note the explicit dependence of the $G_\text{YSR}$ on $E_\text{J}$, which is absent in $G_\text{BCS}$, indicating the quite different nature of these two transport channels.

The red line in Fig.\ \ref{fig:plot2}(d) shows a fit to the spectrum involving a transport channel through the YSR state along with a channel through an empty Bardeen-Cooper-Schrieffer (BCS) gap. The individual densities of states for the YSR state (red) and the BCS gap (blue) are shown in Fig.\ \ref{fig:plot2}(e). In the following, we will assume that these two orbitals have the same decay constant into the vacuum in order to keep the model simple. The fit (red line in Fig.\ \ref{fig:plot2}(d)) reveals that 22\% of the signal (referenced to the normal state conductance $G_\text{N}$ at high bias voltage) is contributed from the YSR state channel and 78\% of the signal comes from the empty BCS gap channel. We are now in a position to compare the experimental data for $G_\text{N}(z)$ with predictions obtained from the above model (Eq.\ \eqref{eq:ysrcoupling} and \eqref{eq:transmission}).

\begin{figure}
\centerline{\includegraphics[width = \columnwidth]{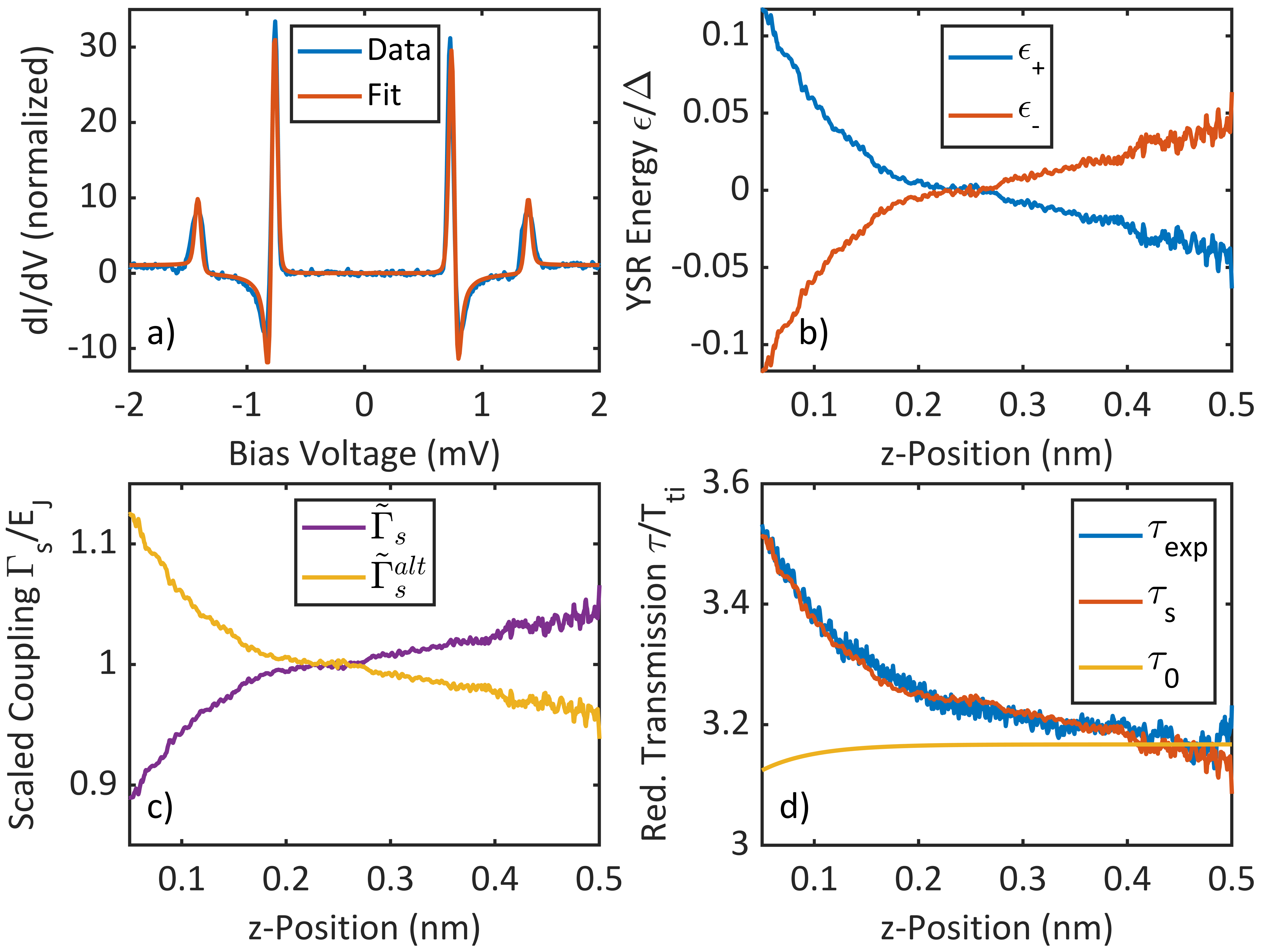}}
\caption{(a) Differential conductance spectrum of a YSR state (blue). The fit (red) considers two channels, one through the YSR state and one through an empty BCS gap. (b) YSR state energy positions as function of tip-sample distance extracted from a data set with high point density along the distance direction. The two branches cross zero energy indicating that they move across the quantum phase transition. (c) Scaled impurity-substrate coupling calculated from the YSR state energies in (b). (d) The reduced transmission $\tau_\text{exp}/\tilde{\Gamma}_\text{t}$ emphasizes the deviation of the experimental data $\tau_\text{exp}$ compared to the model with constant coupling $\tau_0$. We find good agreement with the branch $\tau_\text{s}$, where the impurity-substrate coupling decreases during the tip approach. } \label{fig:analysis1}
\end{figure}

The measured normal state conductance is shown in Fig.\ \ref{fig:analysis2}(a) over about four orders of magnitude. Changes in the exponential behavior are difficult to detect in this graph. A zoom-in to the closer tip-sample distance is shown in Fig.\ \ref{fig:analysis2}(b), where changes in the exponential behavior are most pronounced. Assuming no change in the impurity-substrate coupling, i.\ e.\ $\Gamma_\text{s}=\text{const}$, we calculate the transmission $\tau_0$ from Eq.\ \eqref{eq:transmission}, which is shown as a yellow line in Fig.\ \ref{fig:analysis2}(b). For the decay constant $z_1$, we fit a value of 51.6\,pm. The experimental transmission $\tau_\text{exp}$ clearly increases more than for a constant impurity-substrate coupling. From Eq.\ \eqref{eq:transmission}, we conclude that this can only be explained by a decreasing impurity-substrate coupling, since $G_\text{N}$ is roughly inversly proportional to $\Gamma_\text{s}$. Using the values for $\Gamma_\text{s}$ (cf. Fig.\ \ref{fig:plot2}(c)) in Eq.\ \eqref{eq:transmission}, we plot the resulting transmission $\tau_\text{s}$ as a red line in Fig.\ \ref{fig:analysis2}(b). We find much better agreement with the experimental data $\tau_\text{exp}$ than for the constant impurity-substrate coupling.

Still, the exponential increase of the conductance due to the tunnel coupling masks the agreement. We, therefore, divide all conductance curves by the normalized tunnel coupling $\tilde{\Gamma}_\text{t}=\Gamma_\text{t}/E_\text{J}$ in order to accentuate changes in the exponential dependence. The resulting curves are shown in Fig.\ \ref{fig:analysis2}(d). The deviations from the constant impurity-substrate coupling $\tau_0$ become more obvious now. The experimental data $\tau_\text{exp}$ shows a steady increase as the tip-sample distance decreases significantly deviating from the constant coupling model. The transmission $\tau_\text{s}$ based on the $\Gamma_{s}$ data values extracted from the YSR energies clearly follows the experimental data. We find generally very good agreement, from which we conclude that assigning the negative YSR energy branch in Fig.\ \ref{fig:analysis2}(b) to $\epsilon_+$ is consistent with a decrease of the impurity-substrate coupling as the tip-sample distance decreases and that the system is in the strong scattering regime.

\subsection{Moving across the Quantum Phase Transition}

As another example, we have chosen an intrinsic impurity, for which the YSR state moves across the quantum phase transition, i.\ e.\ the energies cross the zero energy line, when decreasing the tip-sample distance. A differential conductance spectrum with a high point density along the voltage axis (blue) is shown in Fig.\ \ref{fig:analysis1}(a). The YSR states (inner peaks) can be very well seen along with the BCS peaks (outer peaks). The fit (red) again consists of two channels, where 39\% of the signal is contributed from the YSR state channel and 61\% of the signal comes from the empty BCS gap channel.

The extracted YSR state energies are plotted in Fig.\ \ref{fig:analysis1}(b), where the crossing of the energy branches at zero energy is clearly visible. Again, it is \textit{a priori} not possible to decide from which side the system moves across the quantum phase transition. Therefore, we calculate both possibilities for the scaled coupling parameters $\tilde{\Gamma}_\text{s}$ and $\tilde{\Gamma}_\text{s}^\text{alt}$, which are plotted in Fig.\ \ref{fig:analysis1}(c), where one branch increases, while the other branch decreases as function of tip-sample distance.

The excellent agreement between the experiment and the calculation is again accentuated by plotting the transmission curves divided by the normalized exponential tunnel coupling $\tilde{\Gamma}_\text{t}$, which is shown in Fig.\ \ref{fig:analysis1}(d). Comparing $\tau_\text{exp}$ to the transmission $\tau_0$ with constant impurity-substrate coupling $\Gamma_\text{s} = \text{const}$, we find poor agreement. The transmission $\tau_\text{s}$ based on the $\Gamma_\text{s}$ values follows the experimental data very well indicating that the YSR state moves across the quantum phase transition from the strong scattering regime to the weak scattering regime, as we move closer with the tip to the sample. For the tunnel coupling $\tilde{\Gamma}_\text{t}$, we find a decay constant $z_1= 49.15$\,pm. The full conductance dependence can be found in the Supporting Information \cite{supinf}.

\subsection{Increasing Impurity-Substrate Coupling}

\begin{figure}
\centerline{\includegraphics[width = \columnwidth]{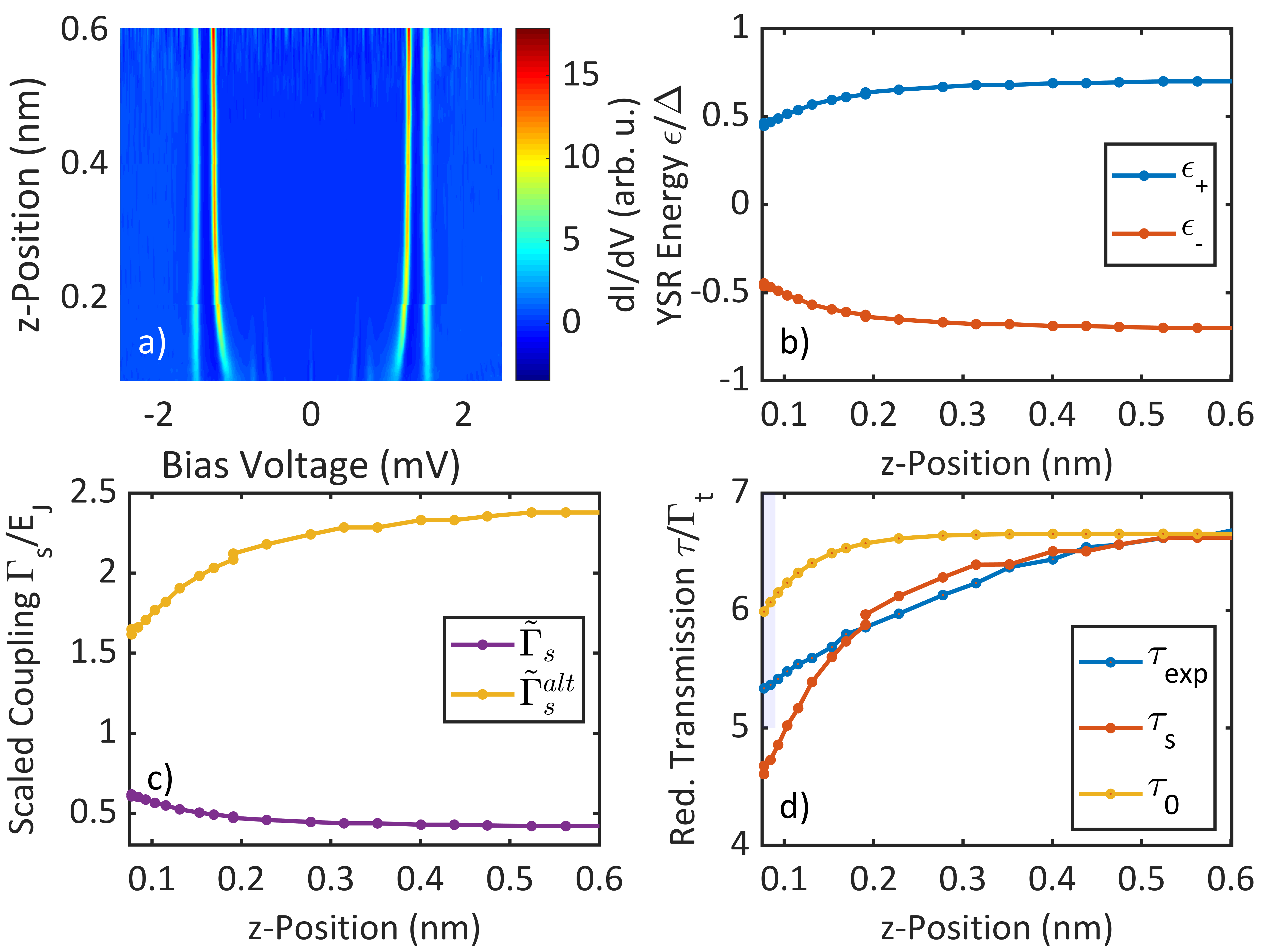}}
\caption{(a) Series of differential conductance spectra (normalized) through an impurity with YSR state measured with a superconducting tip as function of tip-sample distance (z-position). The YSR states move (inner peaks), while the coherence peaks (BCS) do not (outer peaks). At closer distances, higher order phenomena (Josephson effect at zero voltage and multiple Andreev processes near the YSR states) are visible. (b) YSR state energy positions as function of tip-sample distance extracted from the data set in (a). (c) Scaled impurity-substrate coupling calculated from the energies in (b). (d) The reduced transmission $\tau_\text{exp}/\tilde{\Gamma}_\text{t}$ emphasizes the deviation of the experimental data $\tau_\text{exp}$ compared to the model with constant coupling $\tau_0$. We find good agreement with the branch $\tau_\text{s}$, where the impurity-substrate coupling increases during the tip approach. } \label{fig:analysis4}
\end{figure}

As a third example, we found that some of the intrinsic impurities show an increasing impurity-substrate coupling as the tip-sample distance decreases. The image showing the differential conductance spectra as function of applied bias voltage and tip-sample distance (z-position) is plotted in Fig.\ \ref{fig:analysis4}(a). Again the inner peaks are the YSR state and the outer peaks are the BCS coherence peaks. The YSR peaks move towards zero energy as the tip approaches the impurity, while the BCS coherence peaks do not move. The extracted YSR state energies are shown in Fig.\ \ref{fig:analysis4}(b) with both energy branches shown. Using Eq.\ \eqref{eq:ysrcoupling}, we calculate the scaled hopping for both situations $\tilde{\Gamma}_\text{s}$ and $\tilde{\Gamma}_\text{s}^\text{alt}$. Fig.\ \ref{fig:analysis4}(d) shows the transmission curves divided by the normalized exponential tunnel coupling $\tilde{\Gamma}_\text{t}$. Comparing $\tau_\text{exp}$ to the transmission $\tau_0$ with constant impurity-substrate coupling $\Gamma_\text{s} = \text{const}$, we find again poor agreement. We note that the experimental transmission $\tau_\text{exp}$ evolves below the calculated transmission $\tau_0$ (yellow line). This indicates that the impurity-substrate coupling actually increases when approaching the tip to the sample. The transmission $\tau_\text{s}$ based on $\Gamma_\text{s}$ follows the experimental data very well. Here, $\tilde{\Gamma}_\text{s}$ actually increases with decreasing tip-sample distance. For the tunnel coupling $\tilde{\Gamma}_\text{t}$, we find a decay constant $z_1= 52.3$\,pm. The trend clearly indicates that the impurity-substrate coupling increases as we approach with the tip to the sample. This means that the YSR state is in the weak scattering regime. The full conductance dependence can be found in the Supporting Information \cite{supinf}.

\section{Discussion}

\begin{figure}
\centerline{\includegraphics[width = 0.9\columnwidth]{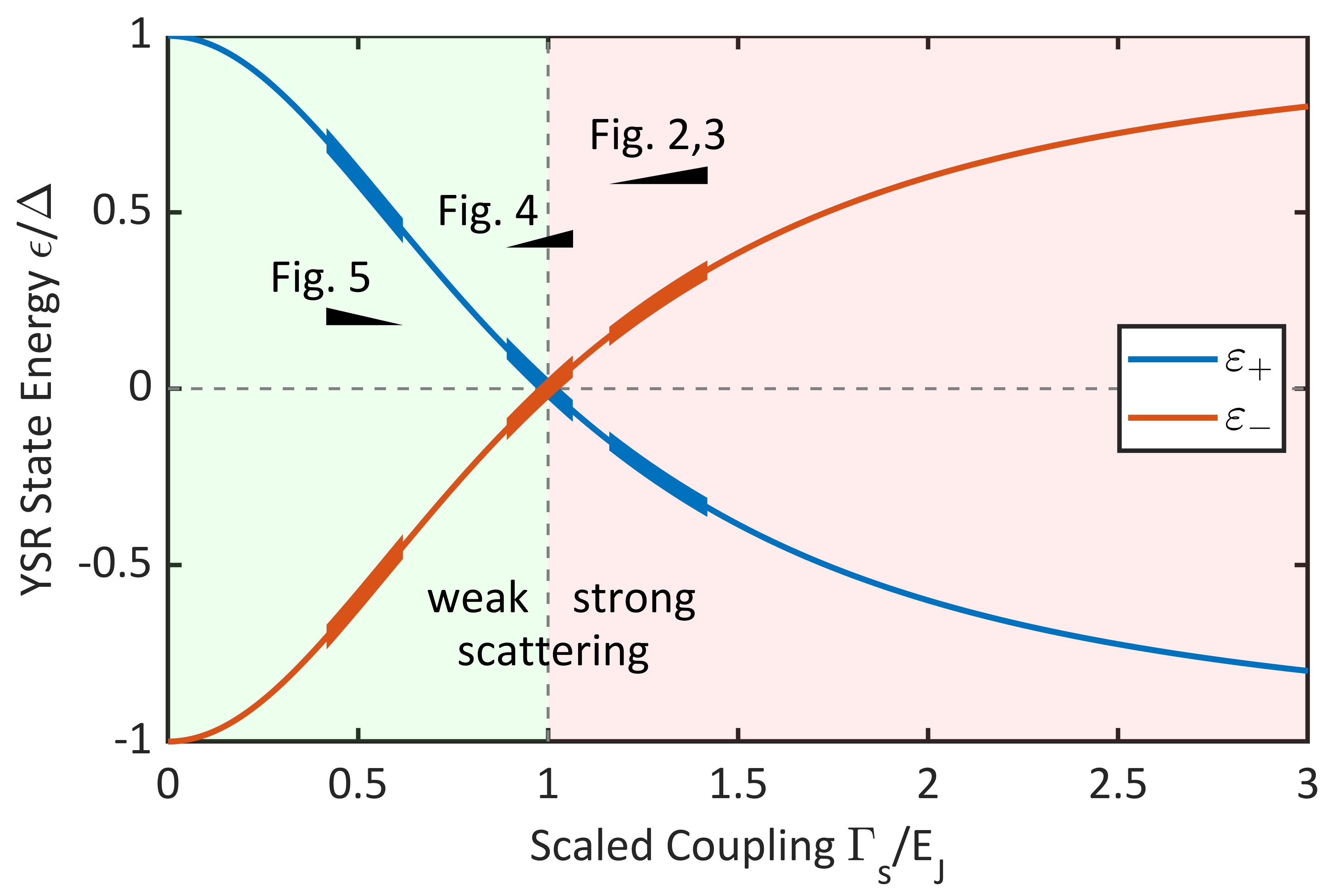}}
\caption{Energies of the YSR state as function of scaled coupling $\Gamma_\text{s}/E_\text{J}$. For small coupling $\Gamma_\text{s}<E_\text{J}$ scattering is weak and the YSR peaks move towards zero for increasing coupling. At zero energy, where $\Gamma_\text{s}=E_\text{J}$ the system undergoes a quantum phase transition into the strong scattering regime. In the strong coupling regime, where $\Gamma_\text{s}>E_\text{J}$, the YSR energies move away from zero as the coupling increases. The range of coupling values in the different data sets are indicated as black wedges with the thinner end indicating a smaller tip-sample distance.} \label{fig:en_vs_gamma}
\end{figure}

Measuring the normal state conductance along with the YSR state energy as a function of tip-sample distance allows us to extract very valuable information, such as an increase or decrease in the impurity-substrate interaction, experimentally without resorting to \textit{ab initio} calculations. The details of the interaction mechanism with the tip and the corresponding change in the impurity-substrate interaction need not be known for an assessment of the coupling regime. We were able to identify on which side of the quantum phase transition the system is for all three examples. In addition, this method can easily be extended to other scenarios presented in the literature \cite{ternes_scanning_2006,brand_electron_2018,farinacci_tuning_2018,malavolti_tunable_2018,kezilebieke_interplay_2018}.

The three examples present different (non-exhaustive) scenarios that can be found when YSR states move in energy as the tip is approaching the impurity. The results are summarized in Fig.\ \ref{fig:en_vs_gamma}, where the energies of the YSR states are plotted as function of the scaled coupling $\Gamma_\text{s}/E_\text{J}$. With the analysis presented above, we can now indicate the coupling range for each example as a black bar labeled by the figure number, where the data set is discussed. Note that the evolution of the YSR state energies and their crossing at the transition point ($\Gamma_\text{s}=E_\text{J}$) nicely illustrates the ambiguity in determining the scattering regime, if the analysis were solely based on the energy position of the YSR state.

The excellent agreement between the measured and calculated normal state transmission $\tau$ clearly identifies the impurity-substrate coupling $\Gamma_\text{s}$ as the dominant energy scale responsible for changing the energy of surface derived YSR states as function of tip-sample distance. This is further corroborated by the conduction channel analysis, showing that the dominant part of the current goes through the empty gap channel, which is unaffected by the magnetic properties of the YSR channel. The intrinsic magnetic properties of the adsorbate remain unchanged at the surface to lowest approximation. This also validates the delicate interplay between the intrinsic magnetic properties of the adsorbate and its interaction with the superconducting host as the responsible mechanism for placing the YSR states inside the gap and even driving them through the quantum phase transition depending on their adsorption site \cite{hatter_magnetic_2015,bauer_microscopic_2013,franke_competition_2011} as well as the tip-sample distance \cite{ternes_scanning_2006,brand_electron_2018,farinacci_tuning_2018,malavolti_tunable_2018,kezilebieke_interplay_2018}.

We find very similar decay constants $z_1$ for the tunnel coupling for all three examples between 49.15\,pm and 52.3\,pm lying within a few percent, which shows that the different examples feature very similar impurities. Interestingly, we have made no explicit assumption about the distance dependence of the impurity-substrate coupling. The impurity-substrate coupling is calculated from the YSR energies and matches well with the conductance change as function of tip-sample distance. This provides a pathway for learning more about the impurity-substrate coupling and the bond strength in particular as function of bond length (i.\ e.\ impurity-substrate distance). Force-distance measurements in a combination of STM with atomic force microscopy (AFM) could provide further insight on the tip-sample interaction as well as the impurity-substrate coupling \cite{ternes_interplay_2011}.

The Anderson impurity model naturally takes into account the impurity-substrate hybridization through an explicit parameter, which is only implicitly contained in the Kondo impurity model. This is important as the surface provides much less constrained boundary conditions for adsorption and relaxation than the much higher coordination requirements in the three-dimensional bulk. Furthermore, the Anderson model enables a more detailed description of the tunneling process \textit{through} the impurity which is largely assumed in the tunneling through YSR states. It provides a direct connection between the impurity-substrate coupling and the normal state conductance, which allows for a direct comparison with experimental data and thus adds deeper understanding of YSR states at surfaces. Although largely equivalent, we, therefore, promote the Anderson impurity model as the preferred model for surface adsorbed impurities.

Putting the mean field approximation of the Anderson impurity model into the context of other existing models for YSR states, in the strong impurity-substrate coupling limit it connects well with the Kondo impurity model \cite{yu_bound_1965,shiba_classical_1968,rusinov_superconductivity_1969,salkola_spectral_1997,flatte_local_1997,%
flatte_local_1997-1,zitko_quantum_2018} and in the weak impurity-substrate coupling limit it connects to the more general Andreev bound states  \cite{pillet_andreev_2010}. Further, it allows us to including correlations (Kondo effect) by going beyond the mean field approximation \cite{anderson_localized_1961,yoshioka_numerical_2000,bauer_microscopic_2013,kezilebieke_interplay_2018,zitko_shiba_2015,kadlecova_quantum_2017,kadlecova_practical_2019,martin-rodero_josephson_2011}, and it extends to a regime, where the impurity-substrate coupling plays a decisive role, i.\ e.\ for impurities at surfaces.

\section{Conclusion}

We have presented direct experimental evidence that the impurity-substrate coupling for adsorbates at surfaces presents an important energy scale largely responsible for the detailed behavior of surface derived YSR states. The behavior of the impurity-substrate coupling (decrease or increase) can be extracted experimentally through the normal state conductance without knowing the details of the actual mechanism and the tip-impurity interaction. It can be used to diagnose, on which side of the quantum phase transition the system is (see Supporting Information \cite{supinf}). Using the mean field approximation of the Anderson impurity model, we were able to make a direct connection between the accompanying change in the YSR state energy and the change in the impurity-substrate coupling for which it provides an explicit parameter. This connection was evidenced through the explicit calculation of the normal state conductance, which is nicely implemented with the Anderson impurity model because it provides a description of tunneling through the impurity directly.

Our results provide a new point of view on the surface induced YSR states and their interactions with the underlying substrate with many possibilities for a deeper understanding provided by the complementary, but more detailed mean field approximation of the Anderson impurity model. The Anderson impurity model provides the basis for moving away from the classical spin model in YSR states and establishing a better link between the experimental observations and the theoretical models, in particular for surface induced YSR states as well as in the presence of the various manifestations of the Kondo effect.

We gratefully acknowledge stimulating discussions with A.\ Kadlecová, T.\ Novotný, M.\ Ternes, and R.\ \v{Z}itko. This work was funded in part by the ERC Consolidator Grant AbsoluteSpin (Grant No.\ 681164) and by the Center for Integrated Quantum Science and Technology (IQ$^\text{ST}$). JA acknowledges funding from the DFG under grant number AN336/11-1. ALY and JCC acknowledge funding from the Spanish MINECO (Grant No. FIS2017-84057-P and FIS2017-84860-R), from the “Mar\'{\i}a de Maeztu” Programme for Units of Excellence in R\&D (MDM-2014-0377).

\strut
\onecolumngrid
\newpage
\begin{center}
\textbf{\large Supplementary Information}
\strut
\vspace{1em}
\end{center}
\setcounter{figure}{0}
\setcounter{table}{0}
\renewcommand{\thefigure}{S\arabic{figure}}
\renewcommand{\thetable}{S\Roman{table}}
\twocolumngrid
\section{Tip and Sample Preparation}

The experiments were carried out in a scanning tunneling microscope (STM) operating at a base temperature of 10\,mK \cite{si_assig_10_2013}. The sample was a V(100) single crystal. To obtain a clean surface, the sample was prepared by multiple cycles of Ar sputtering and subsequent annealing to 700$^{\circ}$C. The tip material was a polycrystalline V wire, which was cut in air and prepared in ultrahigh vacuum by field emission. The cleaned V(100) always features a ($5\times 1$) oxygen reconstruction due to migration of oxygen to the surface during annealing procedure \cite{si_jensen_1982}. Additional impurities include oxygen vacancies, which are most abundant and visible in STM topography image and some carbon atoms \cite{si_koller_strucutre_2001}. Carbon atoms migrate to the surface during slow cool down procedure after annealing, and is invisible in STM topography measurements \cite{si_bischoff_2002}. Impurities with YSR states can be found on the surface, whose origin is not exactly known. However, from the much smaller abundance compared to oxygen vacancies and the still much higher concentration than remnant transition metal impurities, we conclude that the YSR impurities may originate from combination of the most abundant impurities, such as a carbon-oxygen vacancy complex. The fact that most YSR impurities only show one pair of YSR state in-gap indicates a simple spin structure and supports the simple elements component rather than conventional magnetic transition metal elements.

\section{Mean Field Description of the Anderson Impurity Model}

In the Anderson impurity model, the Hamiltonian $H$ describes the coupling of an impurity $H_\text{i}$ to a substrate $H_\text{s}$ by means of an interaction term $H_\text{si}$ acting like a hopping between impurity and substrate
\begin{eqnarray}
    H & = & H_\text{s} + H_\text{i} + H_\text{si}, \\
    H_\text{s} & = & \sum\limits_{k\sigma} \varepsilon_k c_{k\sigma}^{\dagger}c_{k\sigma} - \Delta_\text{s}\sum\limits_k\left(c_{k\uparrow}^{\dagger}c_{k\downarrow}^{\dagger} + h.c.\right),\\
    H_\text{i} & = & \sum\limits_{\sigma} \varepsilon_d n_{d\sigma} + U_d n_{d\uparrow}n_{d\downarrow},
\end{eqnarray}

\begin{eqnarray}
    H_\text{si} & = & \sum\limits_{k\sigma} t_{kd} \left(c_{k\sigma}^{\dagger}c_{d\sigma} + h.c.\right),
\end{eqnarray}
Here, $\sigma$ denotes the spin states up ($\uparrow$) and down ($\downarrow$) and $n_{d\sigma} = c_{d\sigma}^{\dagger}c_{d\sigma}$ is the number operator. The substrate Hamiltonian $H_\text{s}$ describes a Bardeen-Cooper-Schrieffer (BCS) superconductor, which is readily solved within mean field theory. The energy-momentum relation is given by $\varepsilon_k$ and the superconducting gap is given by $\Delta_\text{s}$. Integrating over momentum, we find the Green's function of the substrate in $2\times2$ Nambu space
\begin{equation}
    G_\text{s}(\omega) = n_\text{s}\frac{(\omega +  i\gamma)\sigma_0 - \Delta_\text{s}\sigma_1}{\sqrt{\Delta_\text{s}^2 - (\omega +  i\gamma)^2}},
\end{equation}
with $n_\text{s}$ being the density of states of the substrate at the Fermi level in the normal conducting state and $\gamma$ being a phenomenological broadening parameter \cite{Dynes}. In the impurity Hamiltonian $H_\text{i}$, the four component operator describing the Coulomb repulsion $U_d$ can be reduced to a two component operator by means of a mean field approximation
\begin{equation}
    U_d n_{d\uparrow}n_{d\downarrow} =
    U_d \langle n_{d\downarrow}\rangle n_{d\uparrow} + U_d \langle n_{d\uparrow}\rangle  n_{d\downarrow},
\end{equation}
where $\langle n_{d\sigma}\rangle$ denotes the spin down density in the spin up level and vice versa. The spin densities of states $\langle n_{d\sigma}\rangle$ depend on each other, but they can be calculated selfconsistently using the equation
\begin{equation}
    \langle n_{d\sigma}\rangle = \int\limits_{-\infty}^{\infty} A_{i\sigma}(\omega)f(\omega)d\omega,
\end{equation}
where $A_{i\sigma}(\omega)=\text{Im}G_{i\sigma}(\omega)$ is the spin-resolved spectral function of the impurity Green's function and $f(\omega)$ is the Fermi-Dirac distribution. Within this mean field approximation, we can rewrite the energy levels of the impurity as
\begin{eqnarray}
    H_\text{i} & = & \left(\varepsilon_d + U_d \langle n_{d\downarrow}\rangle \right)n_{d\uparrow} +\left(\varepsilon_d + U_d \langle n_{d\uparrow}\rangle \right)n_{d\downarrow}\nonumber \\
    & = & \left(-E_\text{J} + E_\text{U} \right)n_{d\uparrow} +\left(E_\text{J} + E_\text{U} \right)n_{d\downarrow}.
    \label{eq:Himpmf}
\end{eqnarray}
The second line in Eq.\ \eqref{eq:Himpmf} defines the energy levels in terms of a splitting parameter $E_\text{J}$ and an offset parameter $E_\text{U}$, which are used in the main text. The relation to the first line illustrates the connection to the mean field description of the Anderson impurity model. We should point out that the relation of $E_\text{J}$ and $E_\text{U}$ to $\varepsilon_d$ and $U_d$ is likely more complex, when taking into account correlation effects. The impurity Hamiltonian $H_\text{i}$ describes an effective Zeeman splitting without introducing a quantization axis as in a magnetic field. The impurity spin is quantized with $s=\frac{1}{2}$ and rotates freely.

With this mean field approximation in mind, we can write down the full Hamiltonian in matrix form
\begin{equation}
    H = \begin{pmatrix}
        H_\text{s} & H_\text{si} \\
        H_\text{si}^{\dagger} & H_\text{i}
        \end{pmatrix}
        = \begin{pmatrix}
        H_\text{s} & \hat{t} \\
        \hat{t}^{\dagger} & H_\text{i}
        \end{pmatrix}.
\end{equation}
We assume a $2\times 2$ Nambu space for each entry, such that the interaction term becomes $H_\text{si}=\hat{t} = t\sigma_3$, where $t = t_{kd}$ is a momentum independent hopping term and $\sigma_3$ is a Pauli matrix. The corresponding Green's function can be written as
\begin{equation}
    G(\omega) = \begin{pmatrix}
        \omega - H_\text{s} & \hat{t} \\
        \hat{t}^{\dagger} & \omega - H_\text{i}
        \end{pmatrix}^{-1}
\end{equation}
Focusing on the impurity Green's function $G_\text{i}(\omega) = G_{22}(\omega)$, we find the expression
\begin{equation}
    G_\text{i}(\omega) = \left(\omega\sigma_0 - E_\text{J}\sigma_0 + E_\text{U}\sigma_3 - \Gamma_\text{s}\sigma_3g_\text{sc}(\omega)\sigma_3\right)^{-1},
    \label{eq:Gimp}
\end{equation}
where $\sigma_j$ are the Pauli matrices, $\Gamma_\text{s} = |t|^2 n_\text{s}$ is the impurity-substrate coupling parameter, and $g_\text{sc}$ is the normalized Green's function of the superconducting substrate with $G_\text{s}(\omega) = n_0 g_\text{sc}(\omega)$. Equation \eqref{eq:Gimp} represents the Green's function of the impurity as used in the main text.

\section{Transport through an Impurity}

Tunneling involving a YSR state is typically thought of as transport through the YSR state, i.\ e.\ transport through the impurity at the surface. The tunneling current through an impurity not only involves the tunnel coupling between tip and impurity, but also between the impurity and the substrate. This is best modelled within the Anderson impurity model as it naturally involves the impurity-substrate coupling and has been discussed extensively in the literature. The normal state conductance is typically measured at bias voltages $V$ much larger than the superconducting gaps, i.\ e.\ $eV\gg\Delta_\text{t}+\Delta_\text{s}$, where the influence of the superconducting gaps on the normal state conductance can be neglected. In this approximation, the transmission through the YSR state can be written as \cite{si_cuevas_molecular_2010}
\begin{equation}
    \tau = \frac{4 t^2_\text{ti}n_\text{i} n_\text{t}}{|1-t_\text{ti}^2g_\text{i}g_\text{t}|^2},
\end{equation}
where $n_\text{i}$ is the density of states at the impurity,  $n_\text{t}$ is the density of states in the tip, $g_\text{i}$ is the Green's function of the impurity in the normal conducting state, $g_\text{t}$ is the density of states in the tip in the normal conducting state, and $t^2_\text{ti}$ describes the tunnel hopping between the tip and the impurity. Assuming particle-hole symmetry, i.\ e.\ $E_\text{U}=0$, the Green's functions are
\begin{equation}
    g_\text{i} = \frac{E_\text{J} + i\Gamma_\text{s}}{\Gamma_\text{s}^2+E_\text{J}^2}\quad\text{and}\quad g_\text{t} = i n_\text{t}
\end{equation}
and the density of states $n_\text{i}$ in the impurity is
\begin{equation}
    n_\text{i}= \text{Im}\left(g_\text{i}\right) = \frac{\Gamma_\text{s}}{\Gamma_\text{s}^2+E_\text{J}^2}.
\end{equation}
Further defining $\Gamma_\text{t} = t^2_\text{ti}n_\text{t}$, we can calculate the transmission through the YSR state of the impurity
\begin{equation}
    \tau_\text{YSR}=\frac{4\Gamma_\text{s} \Gamma_\text{t}}{(\Gamma_\text{s} + \Gamma_\text{t})^2+E_\text{J}^2}
    = \frac{4\tilde{\Gamma}_\text{s}\tilde{\Gamma}_\text{t}}{(\tilde{\Gamma}_\text{s}+\tilde{\Gamma}_\text{t})^2+1}
    \label{eq:tauysr}
\end{equation}
In the last expression of Eq.\ \eqref{eq:tauysr}, we relate the impurity-substrate coupling $\Gamma_\text{s}$ to the level splitting $E_\text{J}$, which we assume to be constant, through $\tilde{\Gamma}_\text{s} = \frac{\Gamma_\text{s}}{E_\text{J}}$. Furthermore, we define $\tilde{\Gamma}_\text{t} = \frac{\Gamma_\text{t}}{E_\text{J}}$. This allows us to directly relate the experimentally extracted impurity-substrate coupling in the expression for the YSR state energy with the impurity-substrate coupling in the expression for the channel transmission.

The experimental spectra show pronounced coherence peaks, which are not part of the density of states on the YSR state. We, therefore, assume that a second transport channel going through the impurity, but not through the YSR state, is involved. This means that both transport channels share the same impurity-substrate transport channel. Assuming a conventional BCS-type superconducting gap based on a constant Green's function $g_\text{I} = i/\Gamma_\text{s}$ and density of states $n_\text{I} = 1/\Gamma_\text{s}$ , we find for the normal state transmission
\begin{equation}
    \tau_\text{BCS} = \frac{4\Gamma_\text{s} \Gamma_\text{t}}{(\Gamma_\text{s}+\Gamma_\text{t})^2}
    = \frac{4\tilde{\Gamma}_\text{s}\tilde{\Gamma}_\text{t}}{(\tilde{\Gamma}_\text{s}+\tilde{\Gamma}_\text{t})^2}.
    \label{eq:taubcs}
\end{equation}
Further assuming that the two channels are independent, we can add them to find the total normal state conductance $G_\text{N}$ weighted by the ratio $p$ with which they have been measured
\begin{equation}
    G_\text{N} = G_\text{YSR} + G_\text{BCS} = p \frac{4\tilde{\Gamma}_\text{s}\tilde{\Gamma}_\text{t}}{(\tilde{\Gamma}_\text{s}+\tilde{\Gamma}_\text{t})^2+1}G_0 + (1-p)\frac{4\tilde{\Gamma}_\text{s}\tilde{\Gamma}_\text{t}}{(\tilde{\Gamma}_\text{s}+\tilde{\Gamma}_\text{t})^2}G_0,
\end{equation}
Here, $G_0$ is the quantum of conductance. As can be seen, the conductance does not only carry information about the coupling between the tip and the impurity, but also between the impurity and the substrate. As such, it provides excellent means to confirm the connection between the YSR state energy as well as the impurity-substrate coupling.

\section{Conductance Measurements as Function of Tip-Sample Distance}

\begin{figure}
\centerline{\includegraphics[width = \columnwidth]{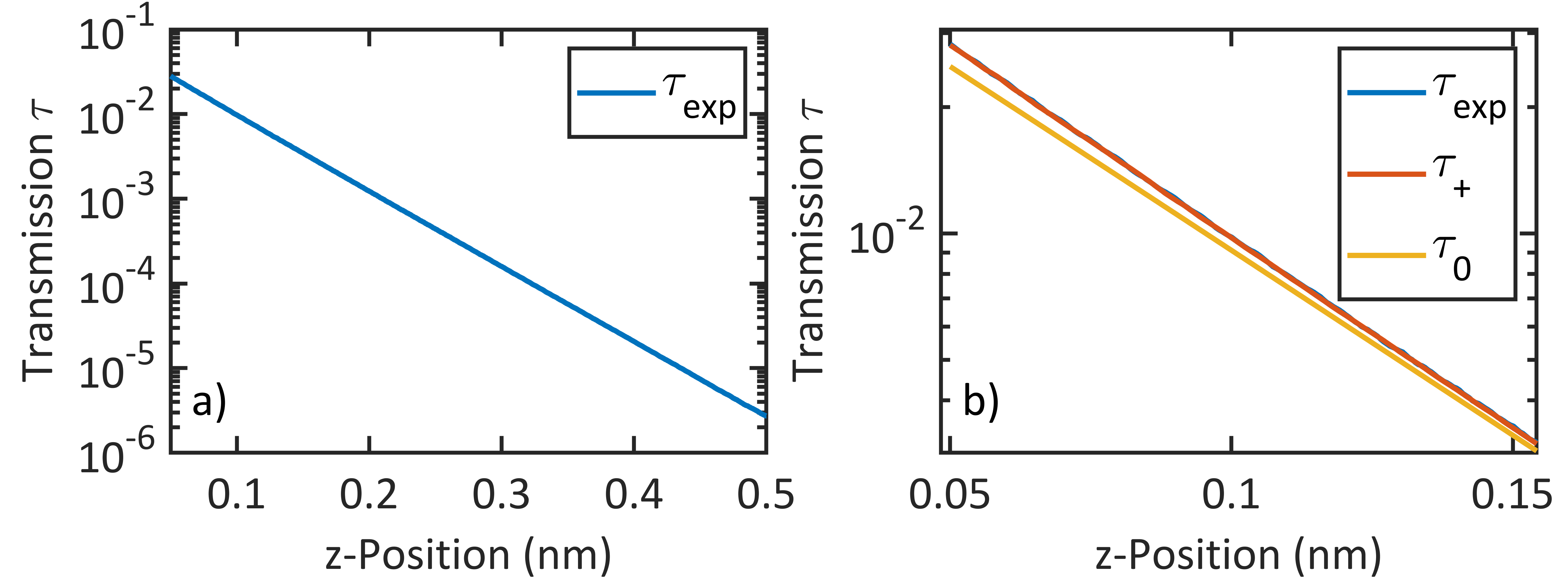}}
\centerline{\includegraphics[width = \columnwidth]{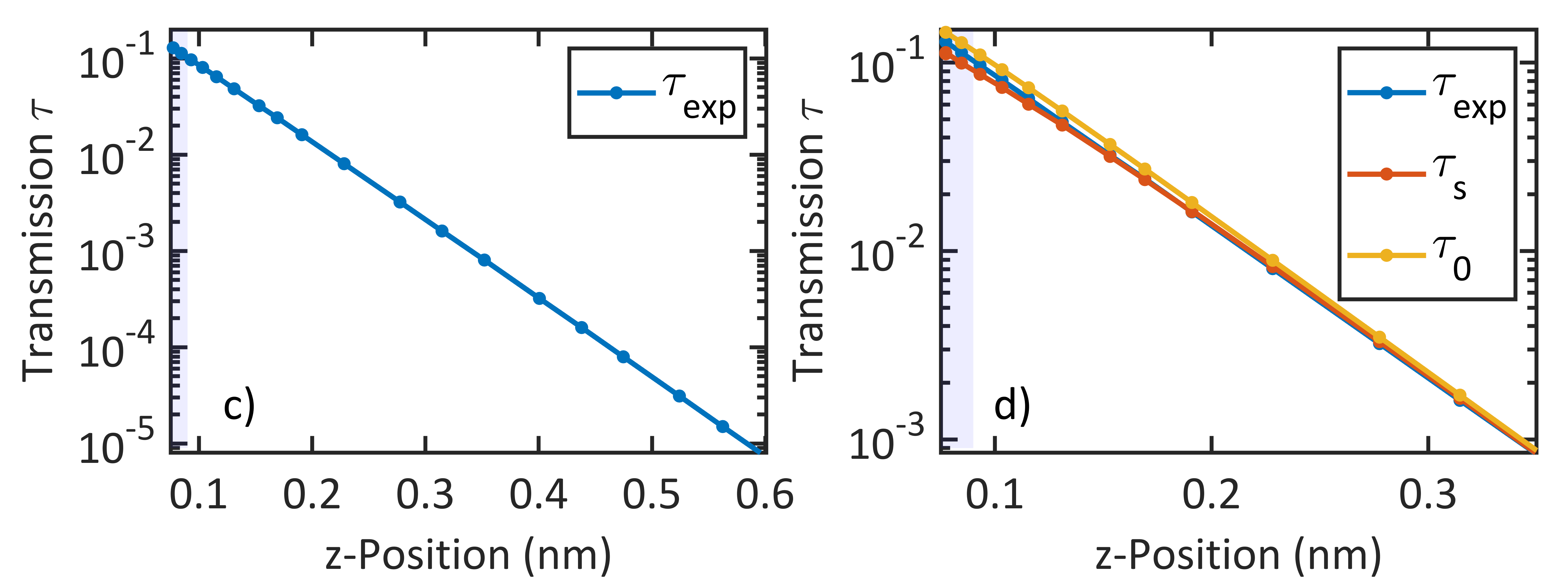}}
\caption{(a) Experimental normal state conductance $G_\text{N}$ as function of tip-sample distance $z$ showing the exponential dependence for the data set presented in Fig.\ 4 of the main text. (b) Zoom-in to the high conductance regime, where the strongest deviations are expected between the experimental data and the theoretical model with a constant impurity-substrate coupling. The calculated transmission based on the extracted impurity-substrate coupling provides a much better agreement even in the logarithmic plot. (c) and (d) is the same as (a) and (b) for the data in Fig.\ 5 of the main text. } \label{fig:fig4and5si}
\end{figure}

For completeness, we show the normal state conductance curves as function of tip-sample distance in Fig.\ \ref{fig:fig4and5si} for the data sets in Figs.\ 4 and 5 of the main text. The exponential dependence of the tunnel coupling dominates the overall behavior, which does not illustrate the comparatively small, but important changes in the impurity-substrate coupling. Therefore, we only show the reduced transmission $\tau/\tilde{\Gamma_\text{t}}$ in the main text.

\section{Experimentally deciding the scattering regime}

If the YSR states are changing their energy $\varepsilon_\pm$, when the tip approaches, the scattering regime can be determined. Both $\varepsilon_\pm$ and the normal state conductance $G_\text{N}$ have to be measured as function of tip-sample distantce $z$. If the conductance $G_\text{N}$ increases more than exponentially as the tip approaches, the impurity-substrate coupling $\Gamma_\text{s}$ decreases. If $G_\text{N}$ increases less than exponentially, $\Gamma_\text{s}$. Whether the scattering regime is weak or strong can be found in Table \ref{tab:regime}.

\begin{table}[t]
\caption{Table to decide, if the system is in the weak or strong scattering regime. Both the YSR state energy $\varepsilon_\pm$ and the normal state conductance $G_\text{N}$ have to be evaluated as function of tip-sample distance $z$.}
\label{tab:regime}
\begin{ruledtabular}
\begin{tabular}{cccccc}
 tip-sample distance decreases & $\Gamma_\text{s}$ increases & $\Gamma_\text{s}$ decreases \\
\hline
$\varepsilon_\pm$ moves towards 0 & weak & strong \\
$\varepsilon_\pm$ moves away from 0 & strong & weak \\

\end{tabular}
\end{ruledtabular}
\end{table}

\section{Fitting the differential conductance}

The differential conductance $dI/dV$ was calculated from the tunneling current
\begin{equation}
I(V) = e\left(\vec{\mathit\Gamma}(V)-\cev{\mathit\Gamma}(V)\right),
\label{eq:iv}
\end{equation}
with the tunneling probability from tip to sample
\begin{widetext}
\begin{equation}
\vec{\mathit\Gamma}(V)=\frac{G_\text{N}}{2e^2}\int\limits^{\infty}_{-\infty}\int\limits^{\infty}_{-\infty}dEdE'(n^e_\text{t}(E-eV)n^e_\text{s}(E')
f(E-eV)[1-f(E')] - n^h_\text{t}(E+eV)n^h_\text{s}(E')f(E+eV)[1-f(E')])P(E-E').
\label{eq:tunprob}
\end{equation}
\end{widetext}
Here, $G_\text{N}$ is the normal state conductance, $f(E)=1/(1+\exp(E/k_\text{B}T))$ is the Fermi function, and $n^{e,h}_\text{t}$, $n^{e,h}_\text{s}$ are the electron ($e$) and the hole ($h$) contributions to the DOS in tip and sample, respectively. For the corresponding Green's functions we use the ($2\times 2$) Nambu space as in the main text. The $P(E)$-function describes the exchange of energy with the environment during the tunneling process and is interpreted as the energy resolution function of the STM \cite{Ast}. The other tunneling direction $\cev{\mathit\Gamma}(V)$ from sample to tip can be obtained by exchanging the Fermi functions in Eq.\ \eqref{eq:tunprob}, i.\ e.\ $f\leftrightarrow (1-f)$.

The DOS in the sample $n_\text{s}$ is modeled by the sum of the DOS of the YSR state $n_\text{YSR}$ and the DOS of the empty BCS gap $n_\text{BCS}$:
\begin{equation}
    n^{e,h}_\text{s} = pn^{e,h}_\text{YSR} + (1-p)n_\text{BCS},
\end{equation}
where $p$ is the relative contributions of the two transport channels to the tunneling current.

The DOS of the YSR state is calculated from the Green's function $G(\omega)$ given in Eq.\ \eqref{eq:SIGYSR}
\begin{equation}
    n^e_\text{YSR}(\omega) = \text{Im} G_{11}(\omega)\quad\text{and}\quad n^h_\text{YSR}(\omega) = \text{Im} G_{22}(\omega).
\end{equation}

The DOS of the empty BCS gap $n_\text{BCS}$ as well as the DOS of the tip $n_\text{t}$ were modeled by the BCS density of states:
\begin{equation}
    n_\text{BCS,t}(\omega) = \Re\left[\frac{\omega+i\gamma_\text{s,t}}{\sqrt{(\omega+i\gamma_\text{s,t})^2-\Delta_\text{s,t}^2}}\right].
\end{equation}
Here, $\gamma_\text{s,t}$ is a phenomenological broadening parameter in sample (s) and tip (t). For the vanadium tip, we find $\Delta_\text{s}=760\,\upmu$eV, $\gamma_\text{s}=\gamma_\text{t} = 3\,\upmu$eV. In order to account for a slight non-BCS shaped tip gap, we use the Maki model for the tip, which introduces a depairing parameter $\zeta_\text{t}$ that leads to an energy dependent tip gap parameter $\Delta_\text{t}$:
\begin{equation}
\Delta_\text{t} = \Delta_\text{t}^\text{BCS} - \zeta_\text{t}\frac{\Delta_\text{t}}{\sqrt{\Delta^2_\text{t} - \omega^2}}.
\end{equation}
with $\Delta_\text{t}^\text{BCS} = 780\,\upmu$eV and $\zeta_\text{t}=0.008$. The sharp YSR state in the sample is a direct probe of the coherence peaks in the tip, which can be directly seen in the tunneling current. The asymmetric shape of the coherence peaks and the falling slope away from the gap result in the typical YSR peak shape in the differential conductance with the characteristic negative differential conductance. Hence, a better fit of the YSR state in the sample can be achieved with a modified coherence peak in the tip. Therefore, we phenomenologically adapt the tip gap by employing the Maki model, which can be rationalized by previously observed differences in the vanadium tip gaps compared to the vanadium bulk gap \cite{si_eltschka_probing_2014,si_eltschka_superconducting_2015,si_jack_nanoscale_2015}.

Differential conductance spectra were recorded with a lock-in amplifier having a modulation amplitude of $20\,\upmu$V and a modulation frequency of $793\,$Hz. The additional broadening in the differential conductance spectra due to the lock-in amplitude is accounted for in the data analysis by means of an additional convolution with a semi-circle shaped resolution function.

\begin{table}[t]
\caption{Fit parameters for the data sets presented in the main text. The parameters $x$ and $y$ are defined in Eq.\ \eqref{eq:xydefs}, $p$ is the contribution of the YSR state to the total normal state conductance, and $\tau$ is the total normal state conductance in units of the quantum of conductance $G_0$.}
\label{tab:parameters}
\begin{ruledtabular}
\begin{tabular}{cccccc}
Set & Figure & $x$ & $y$ & $p$ & $\tau$ \\
\hline
1 & 2,3 & $0.42\pm 0.01$  & $1.36\pm 0.01$ & 22\% & $1.6\times 10^{-4}$ \\
2 & 4   & $0.03\pm 0.006$ & $0.86\pm0.005$ & 39\% & $2.5\times 10^{-2}$ \\
3 & 5   & $0.029\pm0.005$ & $0.39\pm0.005$ & 35\% & $1.6\times 10^{-4}$ \\

\end{tabular}
\end{ruledtabular}
\end{table}

\section{Fitting YSR states within the Anderson impurity model}

In the strong impurity-substrate coupling limit ($\Gamma_\text{s}\gg \Delta$), we find the following Green's function for the YSR states within the Anderson impurity model:
\begin{equation}
    G(\omega) = \frac{\Gamma_\text{s}\omega\sigma_0 + (E_\text{J}\sigma_0+E_\text{U}\sigma_3)\sqrt{\Delta_\text{s}^2-\omega^2} + \Gamma_\text{s}\Delta_\text{s}\sigma_1}{2E_\text{J}\Gamma_\text{s}\omega - (\Gamma_\text{s}^2-E_\text{J}^2+E_\text{U}^2)\sqrt{\Delta_\text{s}^2-\omega^2}},
    \label{eq:SIGYSR}
\end{equation}
It can be easily seen that there are more variables to be determined than independent fit parameters. Therefore, it is not possible to find an unambiguous value for every variable from the fit. We have to reduce the effective number of variables by introducing the following dimensionless parameters:
\begin{equation}
    x = \frac{E_\text{U}}{E_\text{J}};\quad y = \frac{\Gamma_\text{s}}{E_\text{J}};\quad u = \frac{\omega}{\Delta}
    \label{eq:xydefs}
\end{equation}
The resulting Green's function in terms of the parameters $x$, $y$, and $u$ is:
\begin{equation}
    G(u) = \frac{1}{E_\text{J}} \frac{(\sigma_0 + x\sigma_3)\sqrt{1 - u^2} + u y\sigma_0 +y \sigma_1}{2 u y + \sqrt{1 - u^2} (1 - x^2 - y^2)}
\end{equation}
The energy position of the YSR state in this notation is
\begin{equation}
    u_{\pm} = \frac{\varepsilon_{\pm}}{\Delta} = \pm \frac{1-x^2-y^2}{\sqrt{(y^2+(1-x)^2)(y^2+(1+x)^2)}}
\end{equation}
The extracted fit parameters for the data sets presented in the main text are summarized in Table \ref{tab:parameters}. We can see that neglecting the particle-hole asymmetry parameter $E_\text{U}$ will not cause a significant error in the analysis of the main text as $x$ is smaller than one.


\begin{thebibliography}{37}
\expandafter\ifx\csname natexlab\endcsname\relax\def\natexlab#1{#1}\fi
\expandafter\ifx\csname bibnamefont\endcsname\relax
  \def\bibnamefont#1{#1}\fi
\expandafter\ifx\csname bibfnamefont\endcsname\relax
  \def\bibfnamefont#1{#1}\fi
\expandafter\ifx\csname citenamefont\endcsname\relax
  \def\citenamefont#1{#1}\fi
\expandafter\ifx\csname url\endcsname\relax
  \def\url#1{\texttt{#1}}\fi
\expandafter\ifx\csname urlprefix\endcsname\relax\def\urlprefix{URL }\fi
\providecommand{\bibinfo}[2]{#2}
\providecommand{\eprint}[2][]{\url{#2}}

\bibitem[{\citenamefont{Anderson}(1959)}]{anderson_theory_1959}
\bibinfo{author}{\bibfnamefont{P.~W.} \bibnamefont{Anderson}},
  \bibinfo{journal}{J. Phys. Chem. Solids}
  \textbf{\bibinfo{volume}{11}}, \bibinfo{pages}{26} (\bibinfo{year}{1959}).

\bibitem[{\citenamefont{Abrikosov and
  Gor'kov}(1959)}]{abrikosov_superconducting_1959}
\bibinfo{author}{\bibfnamefont{A.~A.} \bibnamefont{Abrikosov}}
  \bibnamefont{and} \bibinfo{author}{\bibfnamefont{L.~P.}
  \bibnamefont{Gor'kov}}, \bibinfo{journal}{JETP} \textbf{\bibinfo{volume}{9}},
  \bibinfo{pages}{220} (\bibinfo{year}{1959}).

\bibitem[{\citenamefont{Kondo}(1964)}]{kondo_resistance_1964}
\bibinfo{author}{\bibfnamefont{J.}~\bibnamefont{Kondo}},
  \bibinfo{journal}{Prog. Theor. Phys.} \textbf{\bibinfo{volume}{32}},
  \bibinfo{pages}{37} (\bibinfo{year}{1964}).

\bibitem[{\citenamefont{Yu}(1965)}]{yu_bound_1965}
\bibinfo{author}{\bibfnamefont{L.}~\bibnamefont{Yu}}, \bibinfo{journal}{Acta
  Phys. Sin.} \textbf{\bibinfo{volume}{21}}, \bibinfo{pages}{75}
  (\bibinfo{year}{1965}).

\bibitem[{\citenamefont{Shiba}(1968)}]{shiba_classical_1968}
\bibinfo{author}{\bibfnamefont{H.}~\bibnamefont{Shiba}},
  \bibinfo{journal}{Prog. Theor. Phys.} \textbf{\bibinfo{volume}{40}},
  \bibinfo{pages}{435} (\bibinfo{year}{1968}).

\bibitem[{\citenamefont{Rusinov}(1969)}]{rusinov_superconductivity_1969}
\bibinfo{author}{\bibfnamefont{A.~I.} \bibnamefont{Rusinov}},
  \bibinfo{journal}{JETP Lett.} \textbf{\bibinfo{volume}{9}},
  \bibinfo{pages}{85} (\bibinfo{year}{1969}).

\bibitem[{\citenamefont{Salkola et~al.}(1997)\citenamefont{Salkola, Balatsky,
  and Schrieffer}}]{salkola_spectral_1997}
\bibinfo{author}{\bibfnamefont{M.~I.} \bibnamefont{Salkola}},
  \bibinfo{author}{\bibfnamefont{A.~V.} \bibnamefont{Balatsky}},
  \bibnamefont{and} \bibinfo{author}{\bibfnamefont{J.~R.}
  \bibnamefont{Schrieffer}}, \bibinfo{journal}{Phys. Rev. B}
  \textbf{\bibinfo{volume}{55}}, \bibinfo{pages}{12648} (\bibinfo{year}{1997}).

\bibitem[{\citenamefont{Flatt{\'e} and
  Byers}(1997{\natexlab{a}})}]{flatte_local_1997}
\bibinfo{author}{\bibfnamefont{M.~E.} \bibnamefont{Flatt{\'e}}}
  \bibnamefont{and} \bibinfo{author}{\bibfnamefont{J.~M.} \bibnamefont{Byers}},
  \bibinfo{journal}{Phys. Rev. B} \textbf{\bibinfo{volume}{56}},
  \bibinfo{pages}{11213} (\bibinfo{year}{1997}).

\bibitem[{\citenamefont{Flatt{\'e} and
  Byers}(1997{\natexlab{b}})}]{flatte_local_1997-1}
\bibinfo{author}{\bibfnamefont{M.~E.} \bibnamefont{Flatt{\'e}}}
  \bibnamefont{and} \bibinfo{author}{\bibfnamefont{J.~M.} \bibnamefont{Byers}},
  \bibinfo{journal}{Phys. Rev. Lett.} \textbf{\bibinfo{volume}{78}},
  \bibinfo{pages}{3761} (\bibinfo{year}{1997}).

\bibitem[{\citenamefont{Schrieffer and Wolff}(1966)}]{schrieffer_relation_1966}
\bibinfo{author}{\bibfnamefont{J.~R.} \bibnamefont{Schrieffer}}
  \bibnamefont{and} \bibinfo{author}{\bibfnamefont{P.~A.} \bibnamefont{Wolff}},
  \bibinfo{journal}{Phys. Rev.} \textbf{\bibinfo{volume}{149}},
  \bibinfo{pages}{491} (\bibinfo{year}{1966}).

\bibitem[{\citenamefont{\v{Z}itko et~al.}(2011)\citenamefont{\v{Z}itko,
  Bodensiek, and Pruschke}}]{zitko_effects_2011}
\bibinfo{author}{\bibfnamefont{R.}~\bibnamefont{\v{Z}itko}},
  \bibinfo{author}{\bibfnamefont{O.}~\bibnamefont{Bodensiek}},
  \bibnamefont{and} \bibinfo{author}{\bibfnamefont{T.}~\bibnamefont{Pruschke}},
  \bibinfo{journal}{Phys. Rev. B} \textbf{\bibinfo{volume}{83}},
  \bibinfo{pages}{054512} (\bibinfo{year}{2011}).

\bibitem[{\citenamefont{\v{Z}itko et~al.}(2015)\citenamefont{\v{Z}itko, Lim,
  L\'{o}pez, and Aguado}}]{zitko_shiba_2015}
\bibinfo{author}{\bibfnamefont{R.}~\bibnamefont{\v{Z}itko}},
  \bibinfo{author}{\bibfnamefont{J.~S.} \bibnamefont{Lim}},
  \bibinfo{author}{\bibfnamefont{R.}~\bibnamefont{L\'{o}pez}}, \bibnamefont{and}
  \bibinfo{author}{\bibfnamefont{R.}~\bibnamefont{Aguado}},
  \bibinfo{journal}{Phys. Rev. B} \textbf{\bibinfo{volume}{91}},
  \bibinfo{pages}{045441} (\bibinfo{year}{2015}).

\bibitem[{\citenamefont{Kadlecov{\'a} et~al.}(2017)\citenamefont{Kadlecov{\'a},
  {\v{Z}}onda, and Novotn{\'y}}}]{kadlecova_quantum_2017}
\bibinfo{author}{\bibfnamefont{A.}~\bibnamefont{Kadlecov{\'a}}},
  \bibinfo{author}{\bibfnamefont{M.}~\bibnamefont{{\v{Z}}onda}},
  \bibnamefont{and}
  \bibinfo{author}{\bibfnamefont{T.}~\bibnamefont{Novotn{\'y}}},
  \bibinfo{journal}{Phys. Rev. B} \textbf{\bibinfo{volume}{95}},
  \bibinfo{pages}{195114} (\bibinfo{year}{2017}).

\bibitem[{\citenamefont{Kadlecov{\'a} et~al.}(2019)\citenamefont{Kadlecov{\'a},
  {\v{Z}}onda, Pokorn{\'y}, and Novotn{\'y}}}]{kadlecova_practical_2019}
\bibinfo{author}{\bibfnamefont{A.}~\bibnamefont{Kadlecov{\'a}}},
  \bibinfo{author}{\bibfnamefont{M.}~\bibnamefont{{\v{Z}}onda}},
  \bibinfo{author}{\bibfnamefont{V.}~\bibnamefont{Pokorn{\'y}}},
  \bibnamefont{and}
  \bibinfo{author}{\bibfnamefont{T.}~\bibnamefont{Novotn{\'y}}},
  \bibinfo{journal}{Phys. Rev. Appl.} \textbf{\bibinfo{volume}{11}},
  \bibinfo{pages}{044094} (\bibinfo{year}{2019}).

\bibitem[{\citenamefont{Mart{\'i}n-Rodero and
  Yeyati}(2011)}]{martin-rodero_josephson_2011}
\bibinfo{author}{\bibfnamefont{A.}~\bibnamefont{Mart{\'i}n-Rodero}}
  \bibnamefont{and} \bibinfo{author}{\bibfnamefont{A.~L.}
  \bibnamefont{Yeyati}}, \bibinfo{journal}{Adv. Phys.}
  \textbf{\bibinfo{volume}{60}}, \bibinfo{pages}{899} (\bibinfo{year}{2011}).

\bibitem[{\citenamefont{Cuevas and Scheer}(2010)}]{cuevas_molecular_2010}
\bibinfo{author}{\bibfnamefont{J.~C.} \bibnamefont{Cuevas}} \bibnamefont{and}
  \bibinfo{author}{\bibfnamefont{E.}~\bibnamefont{Scheer}},
  \emph{\bibinfo{title}{Molecular Electronics: An Introduction to Theory and Experiment}} (\bibinfo{publisher}{World Scientific, Singapore}, \bibinfo{year}{2017}).

\bibitem[{\citenamefont{Ternes et~al.}(2008)\citenamefont{Ternes, Lutz,
  Hirjibehedin, Giessibl, and Heinrich}}]{ternes_force_2008}
\bibinfo{author}{\bibfnamefont{M.}~\bibnamefont{Ternes}},
  \bibinfo{author}{\bibfnamefont{C.~P.} \bibnamefont{Lutz}},
  \bibinfo{author}{\bibfnamefont{C.~F.} \bibnamefont{Hirjibehedin}},
  \bibinfo{author}{\bibfnamefont{F.~J.} \bibnamefont{Giessibl}},
  \bibnamefont{and} \bibinfo{author}{\bibfnamefont{A.~J.}
  \bibnamefont{Heinrich}}, \bibinfo{journal}{Science}
  \textbf{\bibinfo{volume}{319}}, \bibinfo{pages}{1066 }
  (\bibinfo{year}{2008}).

\bibitem[{\citenamefont{Ternes et~al.}(2011)\citenamefont{Ternes, Gonz\'{a}lez,
  Lutz, Hapala, Giessibl, Jel\'{i}nek, and Heinrich}}]{ternes_interplay_2011}
\bibinfo{author}{\bibfnamefont{M.}~\bibnamefont{Ternes}},
  \bibinfo{author}{\bibfnamefont{C.}~\bibnamefont{Gonz\'{a}lez}},
  \bibinfo{author}{\bibfnamefont{C.~P.} \bibnamefont{Lutz}},
  \bibinfo{author}{\bibfnamefont{P.}~\bibnamefont{Hapala}},
  \bibinfo{author}{\bibfnamefont{F.~J.} \bibnamefont{Giessibl}},
  \bibinfo{author}{\bibfnamefont{P.}~\bibnamefont{Jel\'{i}nek}},
  \bibnamefont{and} \bibinfo{author}{\bibfnamefont{A.~J.}
  \bibnamefont{Heinrich}}, \bibinfo{journal}{Phys. Rev. Lett.}
  \textbf{\bibinfo{volume}{106}}, \bibinfo{pages}{016802}
  (\bibinfo{year}{2011}).

\bibitem[{\citenamefont{Ternes}(2006)}]{ternes_scanning_2006}
\bibinfo{author}{\bibfnamefont{M.}~\bibnamefont{Ternes}}, Ph.D. thesis,
  \bibinfo{school}{EPFL} (\bibinfo{year}{2006}).

\bibitem[{\citenamefont{Brand et~al.}(2018)\citenamefont{Brand, Gozdzik,
  N{\'e}el, Lado, Fern{\'a}ndez-Rossier, and Kr{\"o}ger}}]{brand_electron_2018}
\bibinfo{author}{\bibfnamefont{J.}~\bibnamefont{Brand}},
  \bibinfo{author}{\bibfnamefont{S.}~\bibnamefont{Gozdzik}},
  \bibinfo{author}{\bibfnamefont{N.}~\bibnamefont{N{\'e}el}},
  \bibinfo{author}{\bibfnamefont{J.~L.} \bibnamefont{Lado}},
  \bibinfo{author}{\bibfnamefont{J.}~\bibnamefont{Fern{\'a}ndez-Rossier}},
  \bibnamefont{and}
  \bibinfo{author}{\bibfnamefont{J.}~\bibnamefont{Kr{\"o}ger}},
  \bibinfo{journal}{Phys. Rev. B} \textbf{\bibinfo{volume}{97}},
  \bibinfo{pages}{195429} (\bibinfo{year}{2018}).

\bibitem[{\citenamefont{Farinacci et~al.}(2018)\citenamefont{Farinacci, Ahmadi,
  Reecht, Ruby, Bogdanoff, Peters, Heinrich, von Oppen, and
  Franke}}]{farinacci_tuning_2018}
\bibinfo{author}{\bibfnamefont{L.}~\bibnamefont{Farinacci}},
  \bibinfo{author}{\bibfnamefont{G.}~\bibnamefont{Ahmadi}},
  \bibinfo{author}{\bibfnamefont{G.}~\bibnamefont{Reecht}},
  \bibinfo{author}{\bibfnamefont{M.}~\bibnamefont{Ruby}},
  \bibinfo{author}{\bibfnamefont{N.}~\bibnamefont{Bogdanoff}},
  \bibinfo{author}{\bibfnamefont{O.}~\bibnamefont{Peters}},
  \bibinfo{author}{\bibfnamefont{B.~W.} \bibnamefont{Heinrich}},
  \bibinfo{author}{\bibfnamefont{F.}~\bibnamefont{von Oppen}},
  \bibnamefont{and} \bibinfo{author}{\bibfnamefont{K.~J.}
  \bibnamefont{Franke}}, \bibinfo{journal}{Phys. Rev. Lett.}
  \textbf{\bibinfo{volume}{121}}, \bibinfo{pages}{196803}
  (\bibinfo{year}{2018}).

\bibitem[{\citenamefont{Malavolti et~al.}(2018)\citenamefont{Malavolti,
  Briganti, H{\"a}nze, Serrano, Cimatti, McMurtrie, Otero, Ohresser, Totti,
  Mannini et~al.}}]{malavolti_tunable_2018}
\bibinfo{author}{\bibfnamefont{L.}~\bibnamefont{Malavolti}},
  \bibinfo{author}{\bibfnamefont{M.}~\bibnamefont{Briganti}},
  \bibinfo{author}{\bibfnamefont{M.}~\bibnamefont{H{\"a}nze}},
  \bibinfo{author}{\bibfnamefont{G.}~\bibnamefont{Serrano}},
  \bibinfo{author}{\bibfnamefont{I.}~\bibnamefont{Cimatti}},
  \bibinfo{author}{\bibfnamefont{G.}~\bibnamefont{McMurtrie}},
  \bibinfo{author}{\bibfnamefont{E.}~\bibnamefont{Otero}},
  \bibinfo{author}{\bibfnamefont{P.}~\bibnamefont{Ohresser}},
  \bibinfo{author}{\bibfnamefont{F.}~\bibnamefont{Totti}},
  \bibinfo{author}{\bibfnamefont{M.}~\bibnamefont{Mannini}},
  \bibnamefont{et~al.}, \bibinfo{journal}{Nano Lett.}
  \textbf{\bibinfo{volume}{18}}, \bibinfo{pages}{7955} (\bibinfo{year}{2018}).

\bibitem[{\citenamefont{Kezilebieke et~al.}(2018)\citenamefont{Kezilebieke,
  {\v{Z}}itko, Dvorak, Ojanen, and Liljeroth}}]{kezilebieke_interplay_2018}
\bibinfo{author}{\bibfnamefont{S.}~\bibnamefont{Kezilebieke}},
  \bibinfo{author}{\bibfnamefont{R.}~\bibnamefont{{\v{Z}}itko}},
  \bibinfo{author}{\bibfnamefont{M.}~\bibnamefont{Dvorak}},
  \bibinfo{author}{\bibfnamefont{T.}~\bibnamefont{Ojanen}}, \bibnamefont{and}
  \bibinfo{author}{\bibfnamefont{P.}~\bibnamefont{Liljeroth}},
  \bibinfo{journal}{arXiv:1811.11591}  (\bibinfo{year}{2018}).

\bibitem[{\citenamefont{Rodrigo et~al.}(2004)\citenamefont{Rodrigo, Suderow,
  and Vieira}}]{rodrigo_use_2004}
\bibinfo{author}{\bibfnamefont{J.~G.} \bibnamefont{Rodrigo}},
  \bibinfo{author}{\bibfnamefont{H.}~\bibnamefont{Suderow}}, \bibnamefont{and}
  \bibinfo{author}{\bibfnamefont{S.}~\bibnamefont{Vieira}},
  \bibinfo{journal}{Eur. Phys. J. B} \textbf{\bibinfo{volume}{40}},
  \bibinfo{pages}{483} (\bibinfo{year}{2004}).

\bibitem[{\citenamefont{Guillamon et~al.}(2008)\citenamefont{Guillamon,
  Suderow, Vieira, and Rodiere}}]{guillamon_scanning_2008}
\bibinfo{author}{\bibfnamefont{I.}~\bibnamefont{Guillamon}},
  \bibinfo{author}{\bibfnamefont{H.}~\bibnamefont{Suderow}},
  \bibinfo{author}{\bibfnamefont{S.}~\bibnamefont{Vieira}}, \bibnamefont{and}
  \bibinfo{author}{\bibfnamefont{P.}~\bibnamefont{Rodiere}},
  \bibinfo{journal}{Physica C: Supercond. Appl.}
  \textbf{\bibinfo{volume}{468}}, \bibinfo{pages}{537} (\bibinfo{year}{2008}).

\bibitem[{sup()}]{supinf}
\bibinfo{note}{See Supporting Information}.

\bibitem[{\citenamefont{Anderson}(1961)}]{anderson_localized_1961}
\bibinfo{author}{\bibfnamefont{P.~W.} \bibnamefont{Anderson}},
  \bibinfo{journal}{Phys. Rev.} \textbf{\bibinfo{volume}{124}},
  \bibinfo{pages}{41} (\bibinfo{year}{1961}).

\bibitem[{\citenamefont{Yoshioka and Ohashi}(2000)}]{yoshioka_numerical_2000}
\bibinfo{author}{\bibfnamefont{T.}~\bibnamefont{Yoshioka}} \bibnamefont{and}
  \bibinfo{author}{\bibfnamefont{Y.}~\bibnamefont{Ohashi}},
  \bibinfo{journal}{J. Phys. Soc. Japan}
  \textbf{\bibinfo{volume}{69}}, \bibinfo{pages}{1812} (\bibinfo{year}{2000}).

\bibitem[{\citenamefont{Vecino et~al.}(2003)\citenamefont{Vecino,
  Mart{\'i}n-Rodero, and Yeyati}}]{vecino_josephson_2003}
\bibinfo{author}{\bibfnamefont{E.}~\bibnamefont{Vecino}},
  \bibinfo{author}{\bibfnamefont{A.}~\bibnamefont{Mart{\'i}n-Rodero}},
  \bibnamefont{and} \bibinfo{author}{\bibfnamefont{A.~L.}
  \bibnamefont{Yeyati}}, \bibinfo{journal}{Phys. Rev. B}
  \textbf{\bibinfo{volume}{68}}, \bibinfo{pages}{035105}
  (\bibinfo{year}{2003}).

\bibitem[{\citenamefont{Mart{\'i}n-Rodero and
  Yeyati}(2012)}]{martin-rodero_andreev_2012}
\bibinfo{author}{\bibfnamefont{A.}~\bibnamefont{Mart{\'i}n-Rodero}}
  \bibnamefont{and} \bibinfo{author}{\bibfnamefont{A.~L.}
  \bibnamefont{Yeyati}}, \bibinfo{journal}{J. Phys.: Cond.
  Mat.} \textbf{\bibinfo{volume}{24}}, \bibinfo{pages}{385303}
  (\bibinfo{year}{2012}).

\bibitem[{\citenamefont{Dynes et~al.}(1978)\citenamefont{Dynes, Narayanamurti,
  and Garno}}]{dynes_direct_1978}
\bibinfo{author}{\bibfnamefont{R.~C.} \bibnamefont{Dynes}},
  \bibinfo{author}{\bibfnamefont{V.}~\bibnamefont{Narayanamurti}},
  \bibnamefont{and} \bibinfo{author}{\bibfnamefont{J.~P.} \bibnamefont{Garno}},
  \bibinfo{journal}{Phys. Rev. Lett.} \textbf{\bibinfo{volume}{41}},
  \bibinfo{pages}{1509} (\bibinfo{year}{1978}).

\bibitem[{\citenamefont{Bauer et~al.}(2013)\citenamefont{Bauer, Pascual, and
  Franke}}]{bauer_microscopic_2013}
\bibinfo{author}{\bibfnamefont{J.}~\bibnamefont{Bauer}},
  \bibinfo{author}{\bibfnamefont{J.~I.} \bibnamefont{Pascual}},
  \bibnamefont{and} \bibinfo{author}{\bibfnamefont{K.~J.}
  \bibnamefont{Franke}}, \bibinfo{journal}{Phys. Rev. B}
  \textbf{\bibinfo{volume}{87}}, \bibinfo{pages}{075125}
  (\bibinfo{year}{2013}).

\bibitem[{\citenamefont{Cuevas et~al.}(1998)\citenamefont{Cuevas, Levy~Yeyati,
  Mart{\'i}n-Rodero, Rubio~Bollinger, Untiedt, and
  Agra{\"\i}t}}]{cuevas_evolution_1998}
\bibinfo{author}{\bibfnamefont{J.~C.} \bibnamefont{Cuevas}},
  \bibinfo{author}{\bibfnamefont{A.}~\bibnamefont{Levy~Yeyati}},
  \bibinfo{author}{\bibfnamefont{A.}~\bibnamefont{Mart{\'i}n-Rodero}},
  \bibinfo{author}{\bibfnamefont{G.}~\bibnamefont{Rubio~Bollinger}},
  \bibinfo{author}{\bibfnamefont{C.}~\bibnamefont{Untiedt}}, \bibnamefont{and}
  \bibinfo{author}{\bibfnamefont{N.}~\bibnamefont{Agra{\"\i}t}},
  \bibinfo{journal}{Phys. Rev. Lett.} \textbf{\bibinfo{volume}{81}},
  \bibinfo{pages}{2990} (\bibinfo{year}{1998}).

\bibitem[{\citenamefont{Hatter et~al.}(2015)\citenamefont{Hatter, Heinrich,
  Ruby, Pascual, and Franke}}]{hatter_magnetic_2015}
\bibinfo{author}{\bibfnamefont{N.}~\bibnamefont{Hatter}},
  \bibinfo{author}{\bibfnamefont{B.~W.} \bibnamefont{Heinrich}},
  \bibinfo{author}{\bibfnamefont{M.}~\bibnamefont{Ruby}},
  \bibinfo{author}{\bibfnamefont{J.~I.} \bibnamefont{Pascual}},
  \bibnamefont{and} \bibinfo{author}{\bibfnamefont{K.~J.}
  \bibnamefont{Franke}}, \bibinfo{journal}{Nat. Commun.} \textbf{\bibinfo{volume}{6}},
  \bibinfo{pages}{8988} (\bibinfo{year}{2015}).

\bibitem[{\citenamefont{Franke et~al.}(2011)\citenamefont{Franke, Schulze, and
  Pascual}}]{franke_competition_2011}
\bibinfo{author}{\bibfnamefont{K.~J.} \bibnamefont{Franke}},
  \bibinfo{author}{\bibfnamefont{G.}~\bibnamefont{Schulze}}, \bibnamefont{and}
  \bibinfo{author}{\bibfnamefont{J.~I.} \bibnamefont{Pascual}},
  \bibinfo{journal}{Science} \textbf{\bibinfo{volume}{332}},
  \bibinfo{pages}{940 } (\bibinfo{year}{2011}).

\bibitem[{\citenamefont{\v{Z}itko}(2018)}]{zitko_quantum_2018}
\bibinfo{author}{\bibfnamefont{R.}~\bibnamefont{\v{Z}itko}},
  \bibinfo{journal}{Physica B: Cond. Mat.}
  \textbf{\bibinfo{volume}{536}}, \bibinfo{pages}{230} (\bibinfo{year}{2018}).

\bibitem[{\citenamefont{Pillet et~al.}(2010)\citenamefont{Pillet, Quay, Morfin,
  Bena, Yeyati, and Joyez}}]{pillet_andreev_2010}
\bibinfo{author}{\bibfnamefont{J.-D.} \bibnamefont{Pillet}},
  \bibinfo{author}{\bibfnamefont{C.~H.~L.} \bibnamefont{Quay}},
  \bibinfo{author}{\bibfnamefont{P.}~\bibnamefont{Morfin}},
  \bibinfo{author}{\bibfnamefont{C.}~\bibnamefont{Bena}},
  \bibinfo{author}{\bibfnamefont{A.~L.} \bibnamefont{Yeyati}},
  \bibnamefont{and} \bibinfo{author}{\bibfnamefont{P.}~\bibnamefont{Joyez}},
  \bibinfo{journal}{Nat. Phys.} \textbf{\bibinfo{volume}{6}},
  \bibinfo{pages}{965} (\bibinfo{year}{2010}).

\end{thebibliography}

\begin{thebibliography}{1}
    \bibitem{si_assig_10_2013}
        M. Assig, \textit{et al.}, \textit{Rev. Sci. Instr.} \textbf{84}, 033903 (2013).
    \bibitem{si_jensen_1982}
        Jensen, V., Andersen, J. N., Nielsen, H. B., and Adams, D. L., \textit{Surf. Sci.} \textbf{116}, 66 (1982).
    \bibitem{si_koller_strucutre_2001}
        R. Koller, \textit{et al.}, \textit{Surf. Sci.} \textbf{480}, 11 (2001).
    \bibitem{si_bischoff_2002}
        Bischoff, M. M. J. \textit{et al.}, \textit{Surf. Sci.} \textbf{513}, 9 (2002).
    \bibitem{Dynes}
        R. Dynes, \textit{et al.}, \textit{Phys. Rev. Lett.} \textbf{41}, 1509 (1978).
    \bibitem{si_cuevas_molecular_2010}
        {J.~C.} Cuevas and {E.}~{Scheer}, \emph{Molecular Electronics} ({World Scientific}, {2010}).
    \bibitem{Ast}
        Ast, C.~R., \textit{et al.}, {\em Nature Commun.}{ \bf 7}, 13009 (2016).
    \bibitem{si_jack_nanoscale_2015}
        Jäck, B., \textit{et al.}, Appl. Phys. Lett. \textbf{106}, 013109 (2015).
    \bibitem{si_devoret_effect_1990}
        Devoret, M.~H., Esteve, D., Grabert, H., Ingold, G., Pothier, H., and Urbina, C. {\em Phys. Rev. Lett.}{ \bf 64}, 1824 (1990).
    \bibitem{si_averin_incoherent_1990}
        Averin, D.~V., Nazarov, Yu.~V., and Odintsov, A.~A., {\em Physica B: Cond. Mat.} {\bf 165}, 945 (1990).
    \bibitem{si_ingold_cooper-pair_1994}
        Ingold, G., Grabert, H., and Eberhardt, U. {\em Phys. Rev. B}{ \bf 50}, 395 (1994).
    \bibitem{si_McMillan}
        W.~L. McMillan, \textit{Phys. Rev.} \textbf{175}, 537 (1968).
    \bibitem{si_Maki}
        K. Maki, \textit{Prog. Theor. Phys.}, \textbf{32}, 29 (1964).
    \bibitem{si_Shiba}
        H. Shiba, \textit{Progress of Theoretical Physics}, \textbf{40}, 435 (1968).
    \bibitem{si_eltschka_probing_2014}
        M. Eltschka, \textit{et al.}, Nano Lett. \textbf{14}, 7171 (2014).
    \bibitem{si_eltschka_superconducting_2015}
        M. Eltschka, \textit{et al.}, Appl. Phys. Lett. \textbf{107}, 122601 (2015).
\end{thebibliography}
\end{document}